\documentclass[manuscript]{aastex}
\usepackage{multirow}

\shorttitle{Correlations in Horizontal Branch Oscillations and Break Components in XTE J1701-462 and GX 17+2}
\shortauthors{Bu et al.}

\begin{document}


\title{Correlations in Horizontal Branch Oscillations and Break Components in XTE J1701-462 and GX 17+2}


\author{Qing-cui Bu\altaffilmark{1} and Li Chen\altaffilmark{1}}
\affil{Department of Astronomy, Beijing Normal University, Beijing 100875, China}
\email{buqc@mail.bnu.edu.cn, chenli@bnu.edu.cn}

\author{Zhao-sheng Li\altaffilmark{2}}
\affil{School of Physics and State Key Laboratory of Nuclear Physics and Technology, Peking University, Beijing 100871, China}

\author{Jin-lu Qu\altaffilmark{3}}
\affil{Laboratory for Particle Astrophysics, CAS, Beijing 100049, China}
\email{qujl@ihep.ac.cn}

\author{T. M. Belloni\altaffilmark{4}}
\affil{INAF-Osservatorio Astronomico di Brera, Via E, Bianchi 46, I-23807 Merate (LC), Italy}
\email{tomaso.belloni@brera.inaf.it}

\and

\author{Liang Zhang\altaffilmark{1}}
\affil{Department of Astronomy, Beijing Normal University, Beijing 100875, China}




\begin{abstract}
  We studied the horizontal branch oscillations (HBO) and the band-limited components observed in the power spectra of the transient neutron star low-mass X-ray binary XTE J1701-462 and the persistent ``Sco-like'' Z source GX 17+2. These two components were studied based on the state-resolved spectra. We found that the frequencies of XTE J1701-462 lie on the known correlations (WK and PBK), showing consistency with other types of X-ray binaries (black holes, atoll sources and millisecond X-ray pulsars). However, GX 17+2 is shifted from the WK correlation like other typical Z sources. We suggest that the WK/PBK main track forms a boundary which separates persistent sources from transient sources. The characteristic frequencies of break and HBO are independent of accretion rate in both sources, although it depends on spectral models. We also report the energy dependence of the HBO and break frequencies in XTE J1701-462 and how the temporal properties change with spectral state in XTE J1701-462 and GX 17+2. We studied the correlation between rms at the break and the HBO frequency. We suggest that HBO and break components for both sources probably arise from a similar physical mechanism: Comptonization emission from the corona. These two components could be caused by same kind of oscillation in a corona who with uneven density, and they could be generated from different areas of corona. We further suggest that different proportions of the Comptonization component in the total flux cause the different distribution between GX 17+2 and XTE J1701-462 in the $rms_{\rm{break}}$-$rms_{\rm{HBO}}$ diagram.
\end{abstract}





\section{Introduction}

Low-frequency quasi-periodic oscillations (LFQPOs) have been detected in most neutron star low mass X-ray binaries (NS-LMXBs) and black hole low mass X-ray binaries (BH-LMXBs) \citep{van06}. Their centroid frequencies are typically $\sim 1-70$ Hz for NS-LMXBs, while $\sim 0.01-30$ Hz for BH-LMXBs. In LMXBs, the timing behavior has been shown to be strongly connected to the spectral properties \citep{has89, van06, van95, hom01, bel04, hom05, bel05}.

Based on X-ray spectra properties and rapid variability behaviors, bright NS-LMXBs are commonly classified into two sub-types, the Z sources and the atoll sources \citep{has89}. The Z sources have luminosity approaching the Eddington limit ($\sim 0.1-1L_{{\rm{Edd}}}$ or more), while the atoll sources show luminosity of $\sim 0.01-0.5L_{{\rm{Edd}}}$. The Color-Diagram (CD) and Hardness Intensity Diagram (HID) of a typical Z source include three branches, the horizontal branch (HB), the normal branch (NB), and the flare branch (FB). Depending on further details about the shape and orientation of these branches, the Z sources were further classified into the ``Cyg-like" Z sources and the ``Sco-like" Z sources. Three distinct types of LFQPOs were observed in the Fourier power density spectra of Z sources: horizontal branch oscillation (HBO) with centroid frequency $\sim 9.5-60$ Hz, normal branch oscillation (NBO) with centriod frequency $\sim 6-20$ Hz, and flare branch oscillation (FBO) with centroid frequency $\sim 20$ Hz. Atoll sources also show $\sim 20-60$ Hz LFQPOs.

In the case of transient BH-LMXBs, LFQPOs have been sorted into type A, B and C QPOs (see \citet{wij99b, cui00, rem02, cas04, zha13, mot11}). NS-LMXBs and BH-LMXBs have plenty of resemblances in their X-ray power spectra \citep{bar95, hei98, oli98, bar00}, in particular at high frequencies \citep{van94a, van94b, ber98}. \citet{cas05} suggested that the properties of type A, B and C LFQPOs in BH-LMXBs are similar to that of FBOs, NBOs, and HBOs in Z sources, respectively. A universal correlation (known as WK correlation) was found between the centroid frequencies of LFQPOs and break frequencies of band-limited components in BH-LMXBs , accreting millisecond pulsars, and atoll sources \citep{wij99a}. Z sources also showed a tight correlation between HBO and break frequencies, only slightly shifted from the BH/atoll track. \citet{psa99} found the PBK correlation which links the lower kilohertz frequency quasi-periodic oscillations (kHz QPO) to LFQPOs in these systems. The LFQPOs involved in WK/PBK correlations in BH-LMXBs and NS-LMXBs were type C QPOs and HBOs, respectively. The WK and PBK correlations found in these different systems suggest that they probably caused by similar physical mechanism \citep{wag01, rez03}.

The origin of LFQPOs in X-ray binaries has been discussed by several authors \citep{ste98, tit99, tag99, don07, sch06, ing09}. The Lense-Thirring (LT) precession mode, first introduced by \citet{ste98} to explain LFQPOs in NS-LMXB is deemed to be the most promising model in explaining LFQPOs. Despite their unclear origin, LFQPOs show a tight correlation with spectral states, which gives us an empirical way to study the accretion disc around compact objects. They could be a key ingredient in understanding the physical mechanism that modulates the different states. Meanwhile, the WK and PBK correlations suggested that LFQPOs probably have a same origin in these different systems, providing us with a new vision to study the physical mechanism of LFQPOs. Correlations as the PBK and WK have been widely studied in the BH and NS sources. The origin of LFQPOs in NS-LMXBs has been a controversial topic ever since they were discovered \citep{ste98, lin12, li14, alt12, hom12}. The special shifted behavior of Z sources in WK/PBK scheme made the physical mechanism of LFQPOs in NS-LMXBs somewhat more confusing.

XTE J1701-462 is a special transient NS-LMXB source discovered by Rossi X-ray Timing Explorer (RXTE) \citep{rem06}, which has gone through both types of Z behaviour and, at the end of its outburst, became an atoll source \citep{hom10}. This source gives us a great opportunity to study the LFQPO mechanism among different subclasses of individual source on the basis of WK/PBK correlation. As a persistent ``Sco-like'' Z source, GX 17+2 displayed a roughly constant level over a long time scale, unlike XTE J1701-462, which was a transient \citep{wij97, hom02}. However, there is a strong similarity in spectral evolution between GX 17+2 and XTE J1701-462 \citep{lin12, li14}. Comparing these two sources could give us a contrast view in studying the properties of HBOs and break components and investigating the shifted behavior of Z sources in WK/PBK correlations. In this paper, we present a detailed analysis of low-frequency power spectra of XTE J1701-462 and GX 17+2. Data reduction and analysis are described in section 2. We give our results in section 3. In section 4, we discuss physical implications of our results. A short summary is represented in section 5.

\section{Data Reduction and Analysis}

\subsection{Data selection}
We analyzed all 866 observations of XTE J1701-462 obtained by the Proportional Counter Array (PCA) instrument on board RXTE during its 2006-2007 outburst. We chose 63 RXTE observations of GX 17+2 obtained in the period 1999 Oct 3-12 (i.e., MJD 51454-51463) for a comparison study. For the reduction, we used the HEASOFT package version 6.12. Only those data with elevation angle $> 10^{\circ}$, a pointing offset $< 0.01$ and South Atlantic Anomaly exclusion time of 30 min were selected for further analysis.

Type I X-ray bursts were identified and removed from the data. Background-subtracted light curves with 16 s time resolution were constructed from the ``standard 2" mode, using data from Proportional Counter Unit (PCU) number 2 only, without dead-time correction. For both XTE J1701-462 and GX 17+2, we defined two X-ray colors: a soft color (SC) and a hard color (HC), as the ratio of counts rates in the $\sim$ 4.5-7.4 keV/2.9-4.1 keV bands (PCA channels 10-17/6-9) and the $\sim$ 10.2-18.1 keV/7.8-9.8 keV bands (PCA channels 24-43/18-23). Intensity was defined as count rate in the energy range $\sim$ 2.9-18.1 keV ( PCA channel 6-43). The small difference in the energy ranges caused by different epochs were ignored. The colors were used to produce color-color diagrams (CDs) and hardness-intensity diagrams (HIDs). For XTE J1701-462, we divided the data into four time intervals, while all observations of GX 17+2 were combined as one interval. Data selections are listed in Tab. 1: for XTE J1701-462, intervals A and B correspond to the Cyg-like stage and intervals C and D to the Sco-like stage. Interval E is that from GX 17+2. The HIDs of all these 5 intervals are plotted in Fig. 1.

\subsection{Timing analysis}

Standard fast Fourier transform (FFT) techniques \citep{van89, van95} were used to create power density spectra (PDS) from all active PCUs, adopting the rms normalisation from \citet{bel90}. High time resolution single-bit and event mode data were used to create PDS, without background-subtraction or dead-time correction. We studied the properties of LFQPOs (in this case, HBOs) and band-limited components by selecting box regions along the horizontal branch in the HID, accumulating what are known as Sz-resolved spectra. The boxes for each interval are shown in Fig. 1: each box corresponds to $> 1600$ s of data. A rank number Sz was given for boxes to track their positions along Z track \citep{has90}. We set the Sz of the HB/NB vertex as 1 and the leftmost point of the HB as 0, also marked in Fig. 1. The Sz of other boxes are obtained by spline interpolation \citep{die00, lin12, li14}.

For each box, PDS are accumulated from 16s segments and a time bin of $2^{-8}$ s (yielding a frequency range 0.0625-128 Hz), then averaged to obtain a single PDS per box. Since few photons are detected in very high energy bands, the channel range considered is 0-149, corresponding to energy range 2-67 keV. We fitted the PDS with a multi-Lorentzian function \citep{now00, bel02}. PDS were plotted in representation of $rms^{2}$ times frequency ($\nu P_{\nu}$). It takes three to four Lorentzians to fit each PDS: the low frequency noise (LFN, also as known as band-limited component), the HBO and its second and third harmonics. Fig. 2 shows a representative PDS of box 11 from interval A, with best fit Lorentzians. Each Lorentzian, whether used to fit a QPO or band-limited noise, yields three parameters: a centroid frequency ($\nu_0$), a full width at half maximum (FWHM), and an integrated fractional rms. Then the characteristic frequency $\nu_{\rm{c}}$ for a particular component can be calculated by using ${\nu_{{\rm{c}}}} = \sqrt {{\nu_0}^2 + {{(FWHM/2)}^2}}$, as introduced by \citet{bel02}. We will indicate the characteristic frequency of the HBO as $\nu_{\rm{HBO}}$ and that of the band-limited component as $\nu_{\rm{break}}$.

In order to study the energy dependence of HBO and band-limited component, we further divided the PCA channel 0-149 interval into 6 sub-channels. The division of energy channels and their corresponding centroid energies are listed in Tab. 2. The multi-Lorentzian fitting results for all selected boxes are list in Tab. 3.



\section{Results}
\subsection{The WK correlation of XTE J1701-462 and GX 17+2}

In the PDS corresponding to different boxes, for XTE J1701-462, we found $\nu_{\rm{HBO}}$ between $\sim 10$ Hz and $\sim 60$ Hz. These HBOs are all accompanied by $\nu_{\rm{break}}$ between $\sim 2$ Hz and $\sim 12$ Hz. For GX 17+2, $\nu_{\rm{HBO}}$ between $\sim 22$ Hz and $\sim 45$ Hz are found, with $\nu_{\rm{break}}$ between $\sim 2$ Hz and $\sim 5$ Hz. We also consider data from \citet{wij99a} and \citet{alt12}. The corresponding WK correlation is shown in Fig. 3. Our points from XTE J1701-462 lie on the main WK track, between the outer edge of atoll sources and lower edge of Z sources, covering most of the points from the 11 Hz accreting pulsar IGR J17480-2446. Meanwhile, the points from Sco-like stage, with frequency range $\sim$ 30-60 Hz, are blanketed with the points from Cyg-like stage. It is worth noticing that data of XTE J1701-462, BHCs and part of atoll sources are on a straight line. However, our data points for GX 17+2 are distributed in the upper envelop of XTE J1701-462, overlapping the data of typical Z sources, showing a slightly shifted correlation from the main cone. It is worth emphasizing that part of the Z data taken from the literature also belonged to GX 17+2, extracted from earlier observation. Our result for the WK correlation is consistent with what \citet{wij99a} had found in GX 17+2.

\subsection{The shifted behavior of Z sources in the WK correlation }

To study the shifted behavior of Z sources, we further plotted the second harmonic frequencies of HBOs ($\nu_2$) for both sources in the WK correlation. We found that the points of $\nu_2$ from XTE J1701-462 overlap with the points of $\nu_{\rm{HBO}}$ from GX 17+2 and typical Z sources. Furthermore, the points of $\nu_2$ from GX 17+2 are distributed on the upper side of typical Z sources.

\subsection{The PBK correlation of XTE J1701-462 and GX 17+2}

We then tested our identification using the PBK relation using the lower kHz QPO frequency ($\nu_{\rm{low}}$) and $\nu_{\rm{HBO}}$. We didn't measure the kHz QPO frequencies ourselves. Since \citet{san10} already found 14 observations with kHz QPO in XTE J1701-462, we further searched the HBO signal in these observations to study the PBK relation. 7 observations are found with both $\nu_{\rm{low}}$ and $\nu_{\rm{HBO}}$ in the Sco-like stage, with their IDs and frequencies listed in Tab. 4. In this case, we only accumulated one PDS per observation in order to directly use the results from \citet{san10}. For GX 17+2, \citet{lin12} had measured the kHz QPOs and HBOs on Sz-resolved spectra. Since we chose the same data segment with \citet{lin12}, we simply used their data, which also listed in Tab. 4. In Fig. 4, we plot the PBK correlation using the data from Tab. 4, together with the data from \citet{psa99}. The data of GX 17+2 are on the main relation, showing good consistency with atoll sources and part of BHCs. However, as can be seen in the inset of Fig. 4, the data of XTE J1701-462 show a larger spread around the correlation than those of GX 17+2, overlapping with the data of GX 17+2 and atoll sources (4U 1608-52 and 4U 1728-34).

\subsection{The energy dependence of HBO and band-limited components in XTE J1701-462}

In order to test the hypothesis that the band-limited and HBO components have the same origin, we analysed the data from the 6 sub-channels (listed in Tab. 2, see above). In XTE J1701-462, 7 boxes (with 6 of them in Cyg-like stage and 1 in Sco-like stage) were selected out from 4 intervals, judging by the appearance of HBO signals in all 6 sub-channels. We also investigated the energy dependence of the rms amplitude of the components ($rms_{\rm{break}}$ and $rms_{\rm{HBO}}$). When calculating the rms amplitude, we took into account the background contribution to convert power to rms amplitude. We neglected the background contribution for the total energy band since its effect is negligible, but considered its contribution for all the sub-channel bands. The formula is like $rms = \sqrt {\frac{{{\rm{P}}}}{{\rm{S + B}}}} * \frac{{{\rm{S + B}}}}{{\rm{S}}}$, where S and B stand for source and background count rates, and P is the power normalized according to Leahy. We also considered the systematic uncertainty into background as introduced by \citet{jah06}. Fig. 5 shows the rms-energy correlation for these two components. The values of $\nu_{HBO}$ are indicated in the upper left corner of each sub-figure. The trends of $rms_{\rm{break}}$ and $rms_{\rm{HBO}}$ with energy are similar in both Cyg-like and Sco-like stages. Both $rms_{\rm{break}}$ and $rms_{\rm{HBO}}$ increase significantly with photon energy below $\sim 12$ keV, and stay more or less constant after $\sim 12$ keV. We used straight lines to fit the first 5 data points for both components, respectively. The fitting results are listed in Tab. 5, with errors given at $1\sigma$ level. R-squared represents the coefficient of determination, indicating how well the model fits the data. We found that the slopes for the break component (a1) are consistent with each other within the error ($\sim 0.2$) in all 7 sub-figures, while those for HBO component (a2) also have a same value ($\sim 0.1$) within the error. However, for GX 17+2, the detection significance of HBOs in 6 energy bands are not good enough to do the similar work as in XTE J1701-462. We could not give a comparative result in GX 17+2.

\subsection{The Sz dependence of HBO and band-limited components in XTE J1701-462 and GX 17+2}

In Fig. 6, we plot $rms_{\rm{break}}$ and $rms_{\rm{HBO}}$ as a function of Sz for all 5 intervals, to study how the temporal properties change with the different positions in the HID. Our results show that both $rms_{\rm{HBO}}$ and $rms_{\rm{break}}$ show a similar trend in all 5 intervals, i.e. they all decrease with Sz. However, the rms values change more obviously in Cyg-like intervals than in Sco-like intervals. For instance, $rms_{\rm{break}}$ (red crosses) decline from 19\% to 4\% in both Cyg-like intervals. $rms_{\rm{break}}$ changes less in Sco-like intervals, decreasing from 16\% to 8\% in Sco-like interval C and dropping from 9\% to 4\% in Sco-like interval D. The values of $rms_{\rm{HBO}}$ decline from 9\% to 2\% in two Cyg-like intervals, while they drop from 6\% to 1\% in Sco-like interval C. The values of $rms_{\rm{HBO}}$ and $rms_{\rm{break}}$ in GX 17+2 are close to those of Sco-like interval D in XTE J1701-462, that is $rms_{\rm{HBO}}$ decrease from $\sim$ 3\% to 1\% and $rms_{\rm{break}}$ decrease from $\sim$ 6\% to 5\%.

\subsection{The $rms_{\rm{break}}$-$rms_{\rm{HBO}}$ correlation in XTE J1701-462 and GX 17+2}

To further investigate the correlations between HBO and break components in XTE J1701-462 and GX 17+2, we plot the relation between $rms_{\rm{break}}$ and $rms_{\rm{HBO}}$ in Fig. 7a. It is interesting to notice that the points of the HB/NB vertex from XTE J1701-462 are distributed separately from those of the HB, with the points of GX 17+2 lying between them. To make the correlation more clear, we further rebinned the adjacent points in terms of $\Delta rms_{\rm{HBO}} < 0.5$ to reduce errors, and plot the result in Fig. 7b. We did not rebin the data of HB/NB vertex because there are only four of them. We used a power law to fit the data of GX 17+2 and J1701-462 separately (excluding the data of HB/NB vertex because of its different physical origin from HB \citep{lin12}). We obtained $rms_{\rm{break}} =(6.69\pm1.17)*{rms_{\rm{HBO}}}^{(0.46\pm0.04)}$ for XTE J1701-462 and $rms_{\rm{break}} =(5.18\pm0.51)*{rms_{\rm{HBO}}}^{(0.17\pm0.09)}$ for GX 17+2, with $1\sigma$ errors. The difference between the two power law indices is different from zero at 99.5\% confidence level (2.8 sigma).

\section{Discussion}

We studied the HBOs and band-limited components observed in the transient NS-LMXB XTE J1701-462 and the persistent ``Sco-like" Z source GX 17+2. Fig. 3 shows that the data points of XTE J1701-462 are on the main relation of WK schema, lying between atoll sources and the typical persistent Z sources (Cyg X-2, GX 5-1, GX 340+0, Sco X-1), without a significant shift. The data points of GX 17+2 are shifted from the main correlation, overlapping with data of the typical persistent Z sources extracted from \citet{wij99a}. In addition, the data points of the second harmonics of the HBO from XTE J1701-462 overlap with the data of fundamental HBO frequency from GX 17+2. Fig. 4 shows that GX 17+2 is on the main PBK correlation, covering the data of typical persistent Z sources, while XTE J1701-462 lies between typical persistent Z sources and atoll sources. Both $rms_{\rm{break}}$ and $rms_{\rm{HBO}}$ increase with photon energy below $\sim 12$ keV and drop above $\sim 12$ keV in XTE J1701-462. Furthermore, both $rms_{\rm{break}}$ and $rms_{\rm{HBO}}$ decrease along HB in XTE J1701-462 and GX 17+2. We also found that $rms_{\rm{break}}$ increases as a power law with $rms_{\rm{HBO}}$ for both sources, only with different rates of increase.

\subsection{WK/PBK implication in XTE J1701-462 and GX 17+2}

XTE J1701-462 is on the main track of WK relation, together with BHCs, atoll sources, while GX 17+2 is shifted from main track like other persistent Z sources (Cyg X-2, GX 5-1, GX 340+0, Sco X-1 and earlier observation of GX 17+2). \citet{wij99a} suggested that the HBOs observed in shifted Z sources are probably harmonic frequencies. However, according to our results, the HBOs in GX 17+2 and XTE J1701-462 should be the same type of QPO since they have similar frequency and rms values. GX 17+2 is a persistent Sco-like Z source, which shows strong similarities in spectral properties with XTE J1701-462 in its Sco-like stage \citep{lin12,li14}. In Fig. 5, one can also see that the $rms_{\rm{HBO}}$ of both GX 17+2 and XTE J1701-462 decreases along the HB, while the values of $rms_{\rm{HBO}}$ in GX 17+2 are close to those of the Sco-like interval D in XTE J1701-462. All these results suggest the HBOs in GX 17+2 have similar properties with those in XTE J1701-462, or even in other types of sources. The HBOs in GX 17+2 should also be fundamental QPOs, not harmonics. Thus, the shifted Z track in WK correlation should be caused by HBOs in persistent Z sources, not harmonics, even the harmonics of HBOs in XTE J1701-462 overlap with the shifted Z track.

XTE J1701-462 has a more scattered distribution than GX 17+2 in the PBK correlation. This could be caused by the existence of two different types of lower kHz QPOs \citep{psa99}. GX 17+2 is a persistent Z source whose temporal and spectral properties barely change with time, which could explain the consistency with other typical persistent Z sources (Cyg X-2, GX 5-1, GX 340+0 and Sco X-1) in the WK and PBK correlations. XTE J1701-462 and GX 17+2 are both Z sources and exhibit strong similarities in spectral and fast variability properties \citep{lin12,li14}. However, they lie on different tracks in the WK and PBK correlations, which makes the physical mechanism behind these correlations more mysterious.

The WK and PBK correlations between these X-ray sources show that the solid surface, the magnetosphere, obsorption along the line of sight do not affect these rapid variability components \citep{wij99a}. These similarities suggested that the basic frequencies of these systems probably have a same origin and are caused by a similar physical mechanism. Our results show that transient sources like XTE J1701-462, BHCs and 4U 1608-52 are on the main track, while persistent sources like typical Z sources and 4U 1728-34 are shifted from the main track. The persistent and transient sources probably have a similar physical mechanism in producing the basic frequencies, but the difference between these two systems may cause the shifted track in WK/PBK schema.

\subsection{The role of accretion rate $\dot{m}$ and the L-T precession model }

For a given source, mass and spin do not change significantly, and most likely only the mass accretion rate ($\dot{m}$) determines $\nu_{\rm{HBO}}$ and $\nu_{\rm{break}}$. Extensive work has been done to study the spectral properties of GX 17+2 and XTE J1701-462 \citep{lin12,li13,lin09}. These authors suggested that the horizontal branches of both GX 17+2 and XTE J1701-462 be interpreted with the same process acting at constant $\dot{m}$, namely an increase of Comptonization, while a changing $\dot{m}$ led to secular changes between subclasses. Tab. 3 shows that $\nu_{\rm{HBO}}$ and $\nu_{\rm{break}}$ increase along HB in both sources at a constant $\dot{m}$, which suggests the values of $\nu_{\rm{HBO}}$ and $\nu_{\rm{break}}$ may be independent with $\dot{m}$, however, under the assumption of certain spectral models.

Several models have been proposed to study LFQPOs. \citet{ing09} described a promising truncated disc/hot inner flow Lense-Thirring (vertical) model to interpret the characteristic frequencies seen in the broadband PSD of BH and NS. Within this model, the frequencies of LFQPOs are mostly dependent on the inner radius of disc, while the break frequency is set by the viscous timescale at the outer edge of the flow (inner radius of disc). The $\sim$ 1 Hz LFQPOs detected in dipping/eclipsing NS-LMXBs also suggested that LT procession models for LFQPOs are possible in NS-LMXBs \citep{hom12}. The LT precession relation ($\nu \propto R^{-3}$ ) arose in GX 17+2, as introduced by \citet{lin12}. However, \citet{li14} found that $\nu_{\rm{HBO}}$ in XTE J1701-462 is positively correlated with inner disc radius based on using same spectra model with GX 17+2, which makes the LT model as an impossible explanation for the HBOs in XTE J1701-462. They further proposed that HBOs come from a corona. In Fig. 3, we show that part of data from XTE J1701-462 overlap with IGR J17480-2446 (green dot), which was assumed to be a inappropriate candidate for LT precession \citep{alt12}. This makes LT precession a controversial model to explain the HBOs in XTE J1701-462 or even in other Z sources. In general, the physical mechanism in WK/PBK correlations can not be well interpreted with any model so far.

\subsection{The physical implication of HBO and break components}

Tab. 5 shows $rms_{\rm{break}}$ increasing linearly with energy below $\sim$ 12 keV at a constant slope (a=0.200$\pm$0.004) for XTE J1701-462 in both Cyg-like and Sco-like stages, while $rms_{\rm{HBO}}$ also shows a same trend with energy at a constant slope (a=0.100$\pm$0.001). Both $rms_{\rm{break}}$ and $rms_{\rm{HBO}}$ stay stable when energy above $\sim$ 12 keV, which is similar to what has been seen for the kHz QPOs \citep{ber96, men01}. In addition, Fig. 6 shows that both $rms_{\rm{break}}$ and $rms_{\rm{HBO}}$ decrease along HB and Fig. 7 also shows that these two components correlate with each other. These above results suggest that break and HBO components may come from a similar physical mechanism. Break and HBO signals reach the strongest strength at $\sim$ 12 keV indicates that they come from higher energy photons, most likely from non-thermal emission from corona, since thermal emission from disk couldn't produce such high energy photons. \citet{li14} also suggested that the HBO components in XTE J1701-462 are possibly generated from Comptonization emission in a corona, based on the spectral fitting results. Here, we further suggest that Comptonization emission in the corona contributes to both break and HBO components in XTE J1701-462. These two components could be caused by the same kind of oscillation in a corona with uneven density, and they could be produced in different area of corona. If break and HBO components come from different areas, they could differ in oscillation frequencies.

GX 17+2 and XTE J1701-462 are Z sources, and the spectral evolution in GX 17+2 was very similar to XTE J1701-462 in its Sco-like stage \citep{lin12}. Our results in Fig. 6 also shows that these two sources have similar variability properties in Sco-like stage. Both $rms_{\rm{HBO}}$ and $rms_{\rm{break}}$ decrease with increased source intensity, with hardness decreasing at the same time. Their power spectra resemble with each other, and they have similar HBO frequencies. In particular, the variations of $rms_{\rm{HBO}}$ and $rms_{\rm{break}}$ in interval E (GX 17+2) are similar with that of interval D (Sco-like stage in XTE J1701-462). These evidences indicate that break and HBO components in GX 17+2 probably have a similar origin as in XTE J1701-462.

Fig. 7b shows that $rms_{\rm{HBO}}$ has a power-law relation with $rms_{\rm{break}}$ for both sources. The data of GX 17+2 lay between the points of HB and the HB/NB vertex of XTE J1701-462. The spectral properties of GX 17+2 are found to be very similar to those of XTE J1701-462 in its Sco-like stage: the HB is associated with Comptonization of disc emission while at the HB/NB vertex the disc assumes a slim disc solution \citep{lin12,li14}. For XTE J1701-462, the different distribution between HB and HB/NB vertex could be caused by the sudden reduction of Comptonization emission when approaching the HB/NB vertex. The connection of non-thermal emission and HBO was found in XTE J1701-462 and GX 17+2: the strength of Comptonization emission decreases when the HBO frequencies increase from upturn to HB, while the HBO signal disappears when Comptonization emission become undetectable on the NB. As shown in Fig. 6, for XTE J1701-462, Cyg-like intervals generally have larger values of rms (break and HBO) than Sco-like intervals, while the range of rms of GX 17+2 is consistent with the Sco-like interval D of XTE J1701-462. We suggest this could be caused by different proportions of Comptonization component in total flux: the proportion of comptonization component in Cyg-like stage is higher than in Sco-like stage which causes higher rms in Cyg-like stage. The proportion of comptonization component in Sco-like stage (GX 17+2) is higher than at the HB/NB vertex, which causes a lowest rms to take place at the HB/NB vertex. Thus, the three distinguishable tracks contributed by HB, GX 17+2 and HB/NB vertex could be caused by different proportions of Comptonization in the total flux.

\subsection{The accretion disc instability model}

The transient properties of some sources can be interpreted by the disc instability model \citep{jea01}, which assumes the accretion disc gains mass from the mass donor at a constant mass transfer rate. The accretion rate is normally lower than the mass supply rate in the disc, thus mass is accumulated in the disc. When the accumulated mass exceeds a certain critical value, a sudden increase of accretion rate results in an outburst. By contrast, the persistent systems have a relatively stable accretion disc. The accretion disc tends to be more unstable in transient sources when they have an outburst, which would affect the oscillation in corona. This could be a reason why transient sources have a wider variation in $\nu_{\rm{HBO}}$ and $\nu_{\rm{break}}$ in WK/PBK schema if they are assumed to arise from different areas of corona. The persistent sources shifted from main track could be caused by the size of corona in persistent sources is different from transient sources. All these suggestions still need further study.

\section{summary}

In this article, we studied the HBOs and band-limited components observed in the unique transient Z source XTE J1701-462 and the persistent ``Sco-like" Z source GX 17+2. Our results show that the WK/PBK main track forms a boundary which separates persistent sources from transient sources. The shifted Z track is probably not caused by by a misidentification of harmonics. The characteristic frequencies are independent of accretion rate in both GX 17+2 and XTE J1701-462, under the assumption of certain spectral models. We suggest that the HBO and break components from both sources probably come from a similar physical mechanism: Comptonization emission in the corona. These two components could be caused by same type of oscillation in corona with uneven density, and they could be produced in different area of corona. We suggest it is the different proportions of the Comptonization component in total flux that cause the difference distribution between GX 17+2 and XTE J1701-462 in $rms_{\rm{break}}$-$rms_{\rm{HBO}}$ diagram.

We are grateful to Y. N. Wang, H. Q. Gao, Z. Zhang for data reduction and theory suggestions. This research has made use of data obtained from the High Energy Astrophysics Science Achieve Research Center (HEASARC) provided by NASA Goddard Space Flight Center. This work is supported by the National Basic Research program of China 973 Program 2009CB824800, the National Natural Science Foundation of China (11173024) and the Fundamental Research Funds for the Central Universities, the Strategic Priority Research Program on Space Science, the Chinese Academy of Sciences, Grant No. XDA04010300. T. M. Belloni acknowledges support from INAF PRIN 2012-6.

\clearpage



\begin{figure}
\includegraphics[scale=.40]{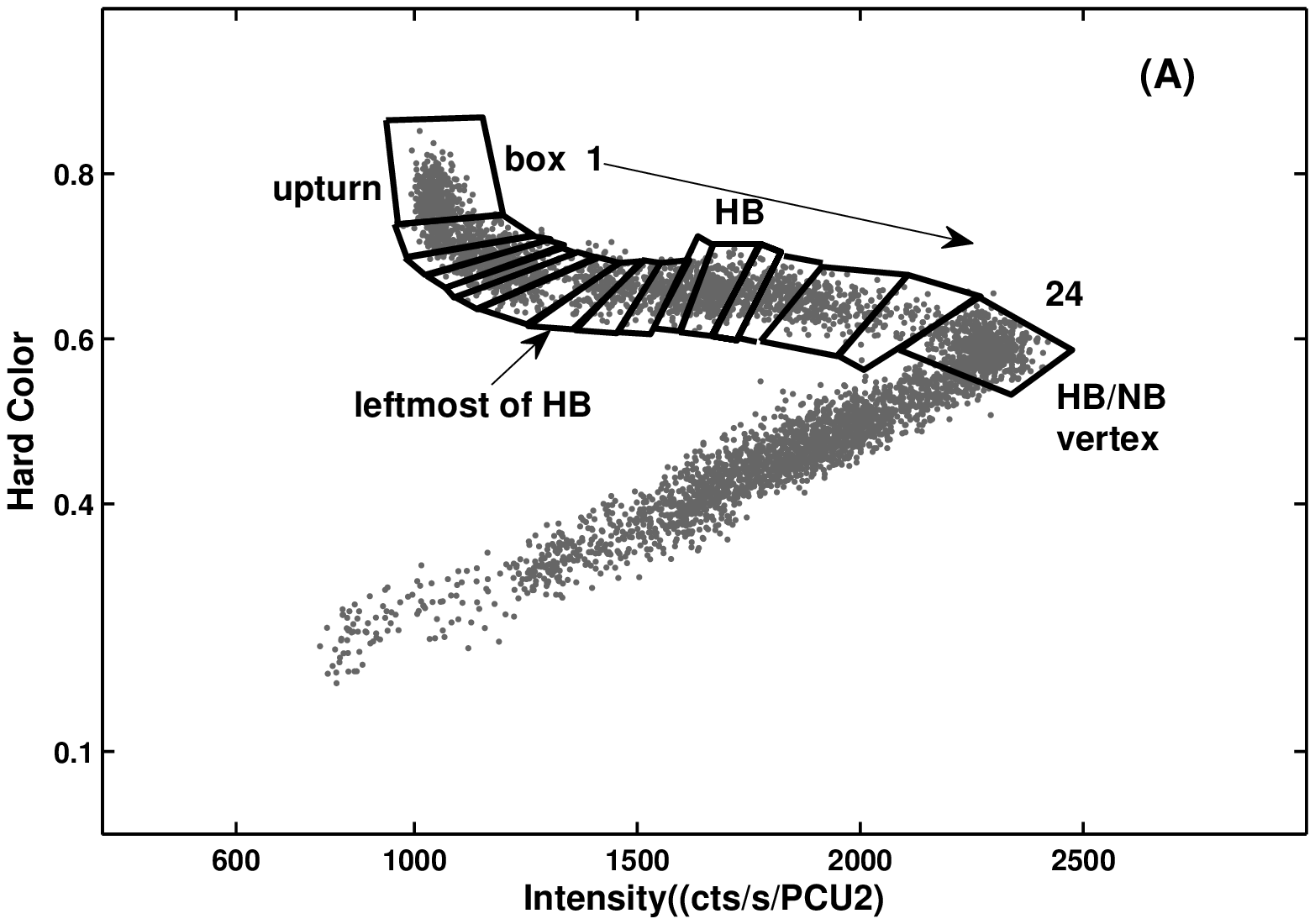}
\includegraphics[scale=.40]{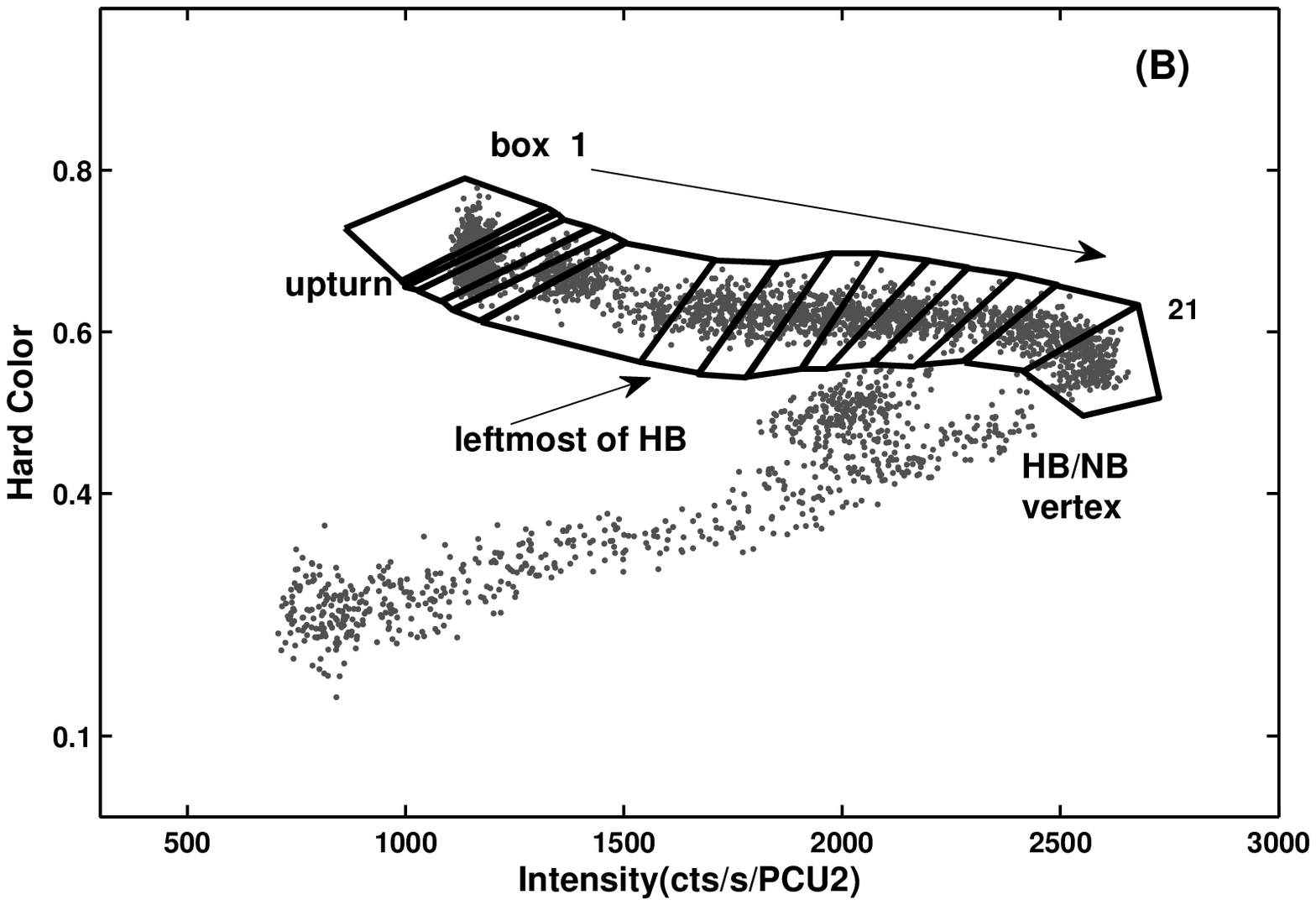} \\
\includegraphics[scale=.40]{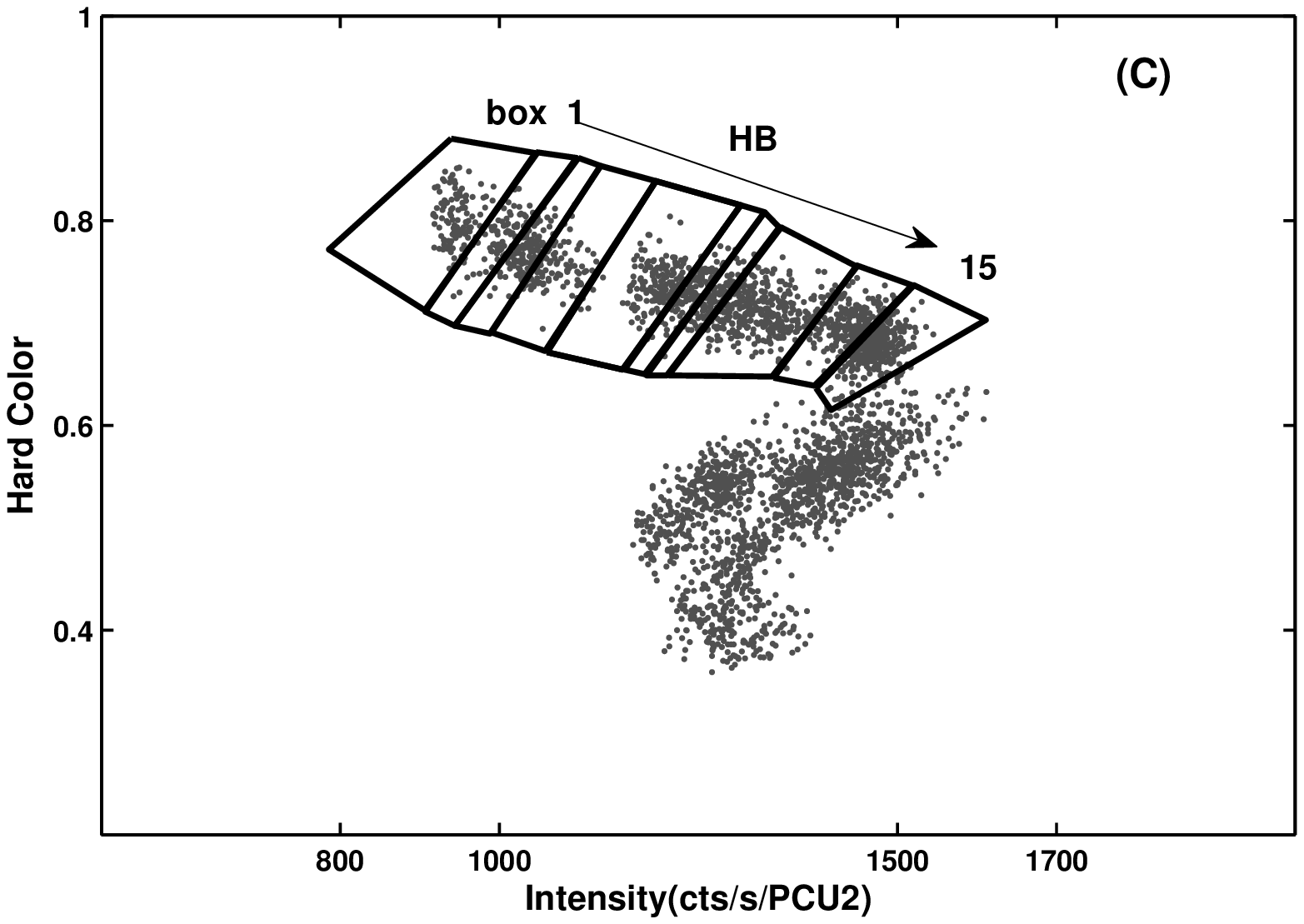}
\includegraphics[scale=.40]{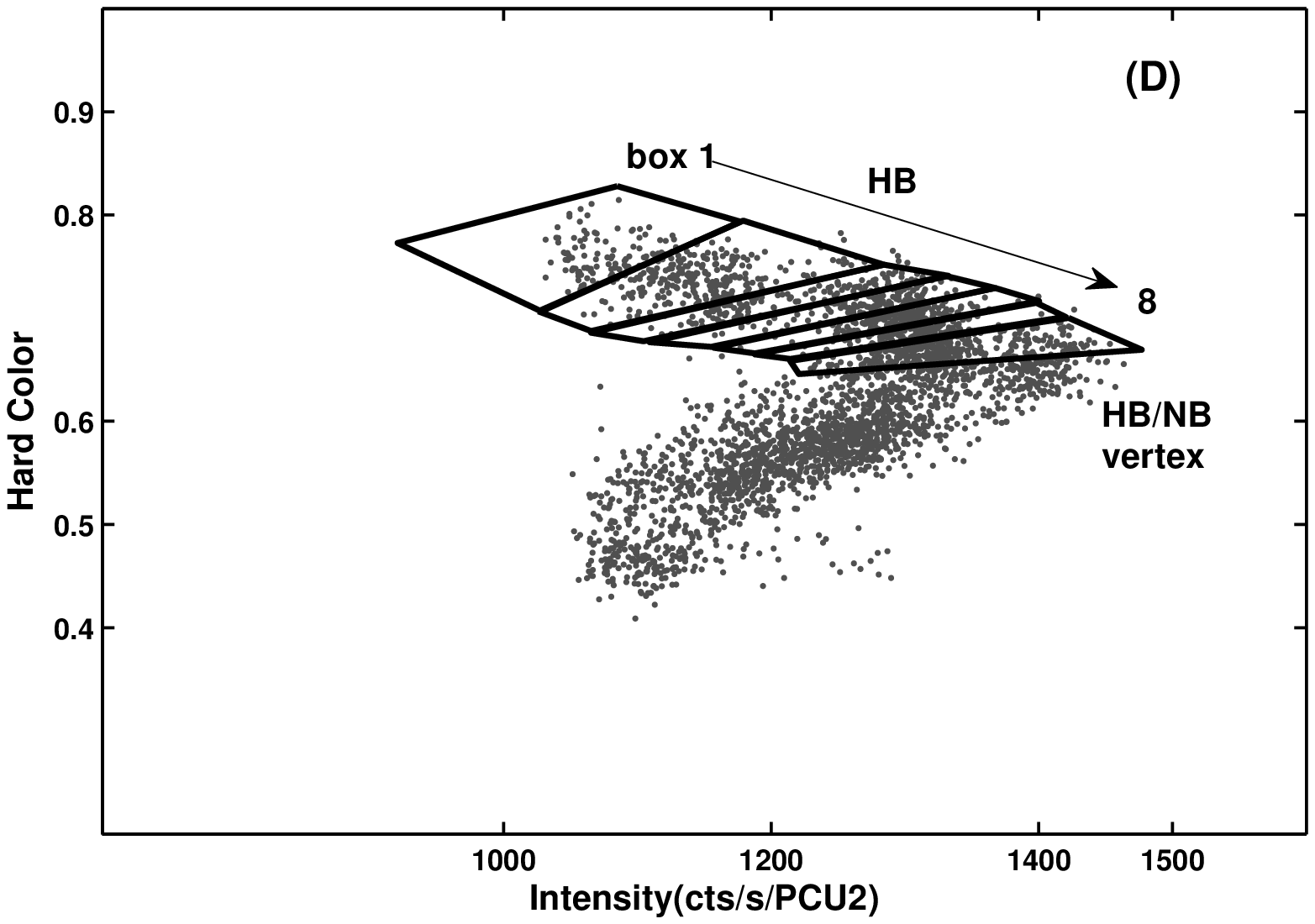} \\
\includegraphics[scale=.40]{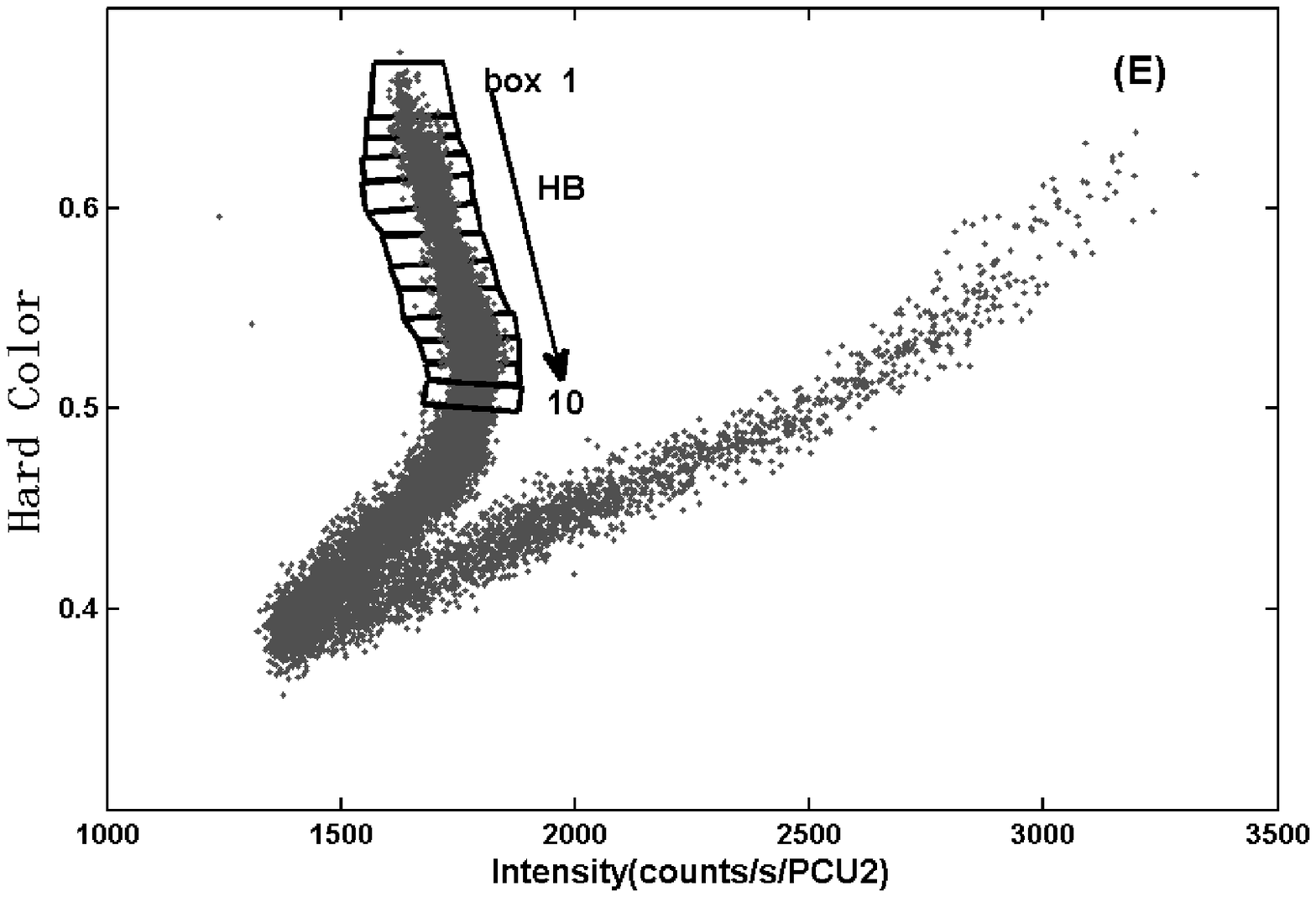}
\caption{Hardness-intensity diagrams and box division of for the five intervals listed in Tab. 1, with their interval number indicated in each panel. A-D are extracted from XTE J1701-462 and E is extracted from GX 17+2. Each dot represents 16 s of background-subtracted data from PCU 2. See more details in the text.}
\label{fig1}
\end{figure}

\clearpage


\begin{figure}
\centering
\epsscale{.8}
\plotone{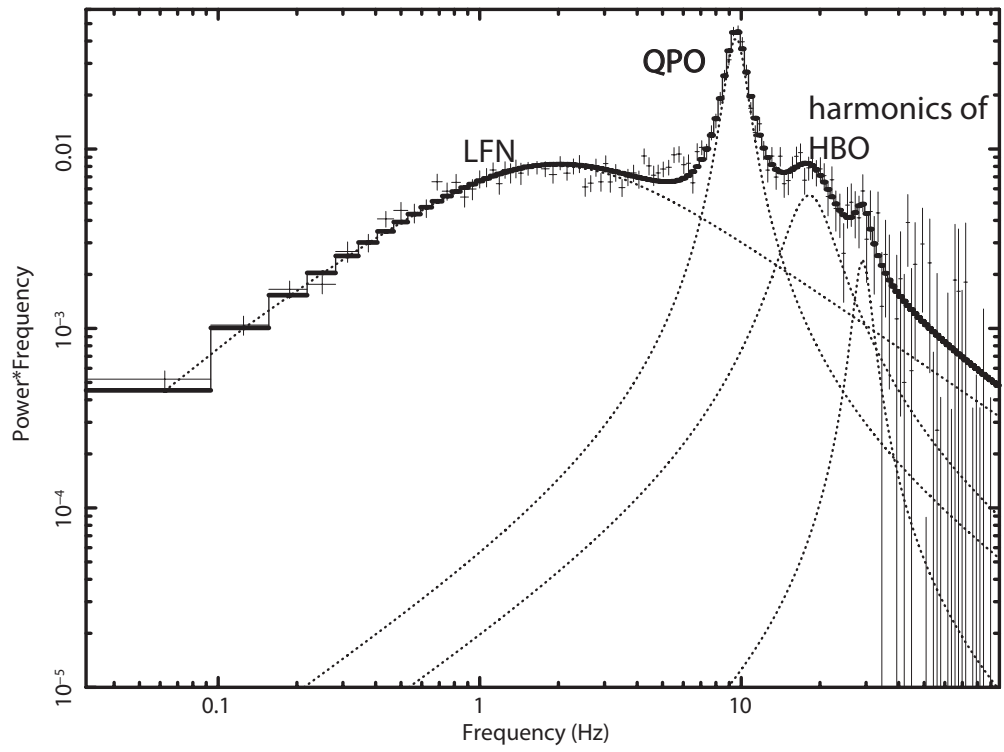}
\caption{Typical power density spectrum of box 11 in interval A of XTE J1701-462, together with the different best fit Lorentzian components.}
\label{fig2}
\end{figure}

\clearpage
\begin{figure}
\centering
\plotone{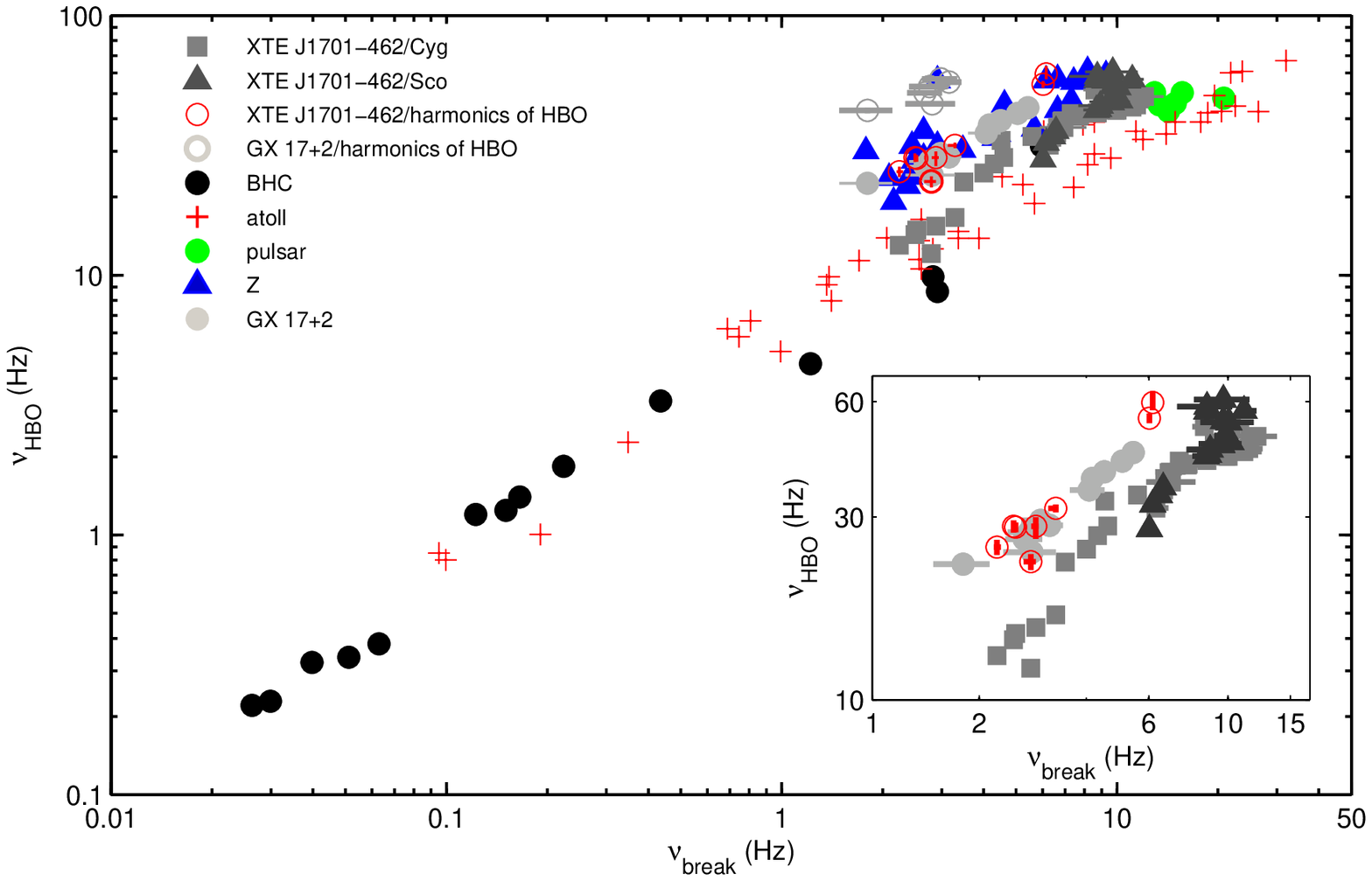}
\caption{The WK correlation invovled with XTE J1701-462 and GX 17+2. The inset panel at the lower right shows a zoomed-in version of our data with error bars.}
\label{fig3}
\end{figure}

\clearpage
\begin{figure}
\centering
\plotone{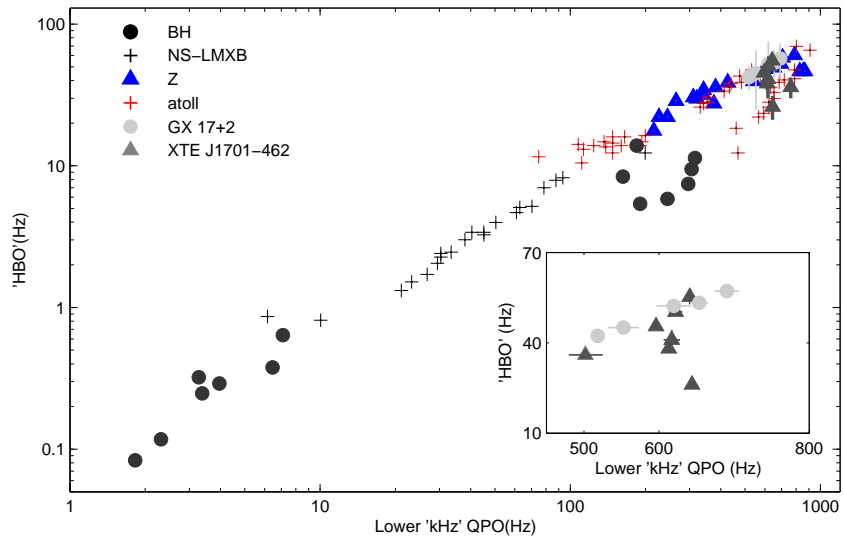}
\caption{The PBK relation included XTE J1701-462 and GX 17+2. The inset panel at the lower right shows a zoomed-in version of our data with error bars.}
\label{fig4}
\end{figure}

\clearpage

\begin{figure}
\includegraphics[scale=.35]{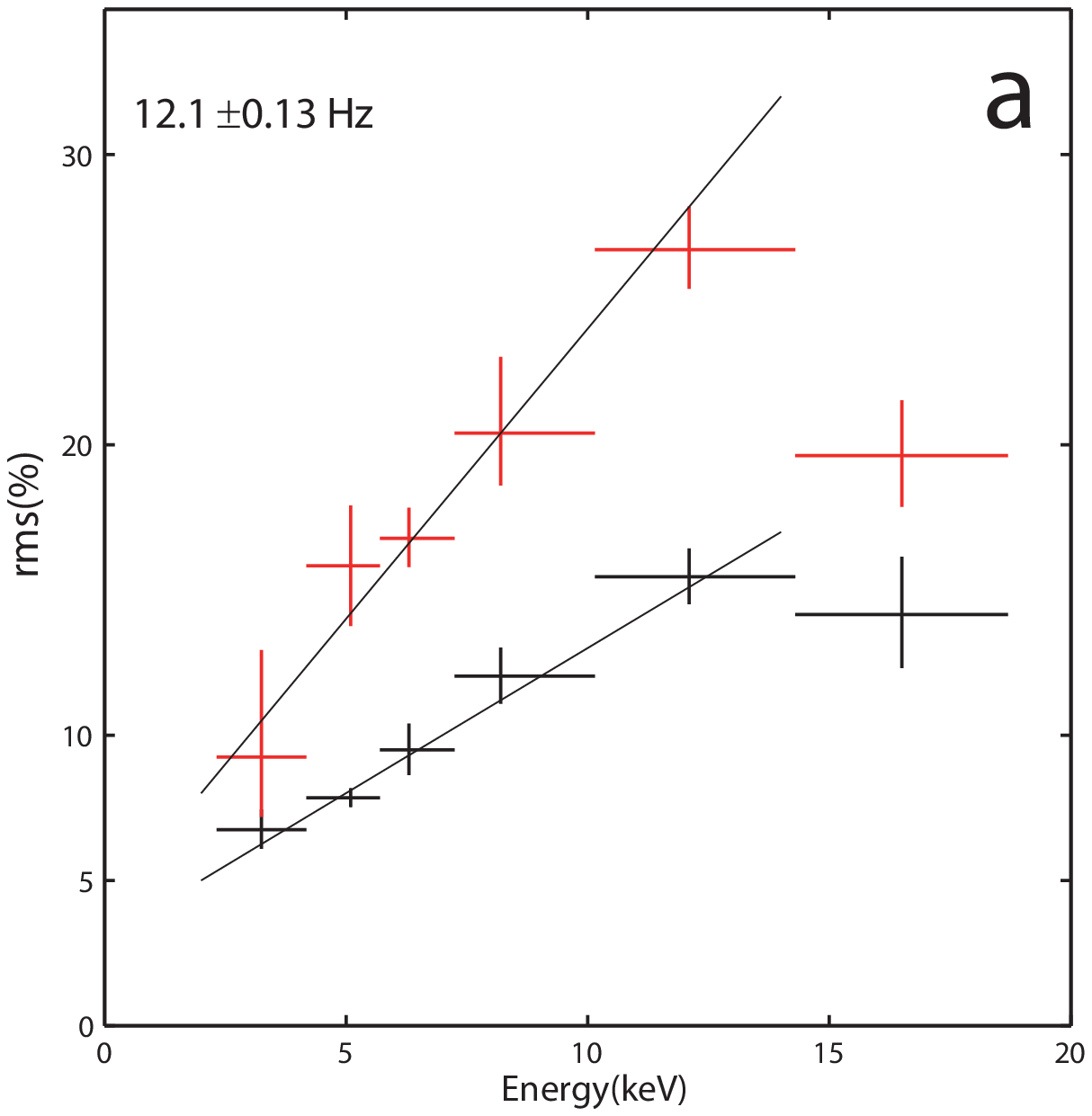}
\includegraphics[scale=.35]{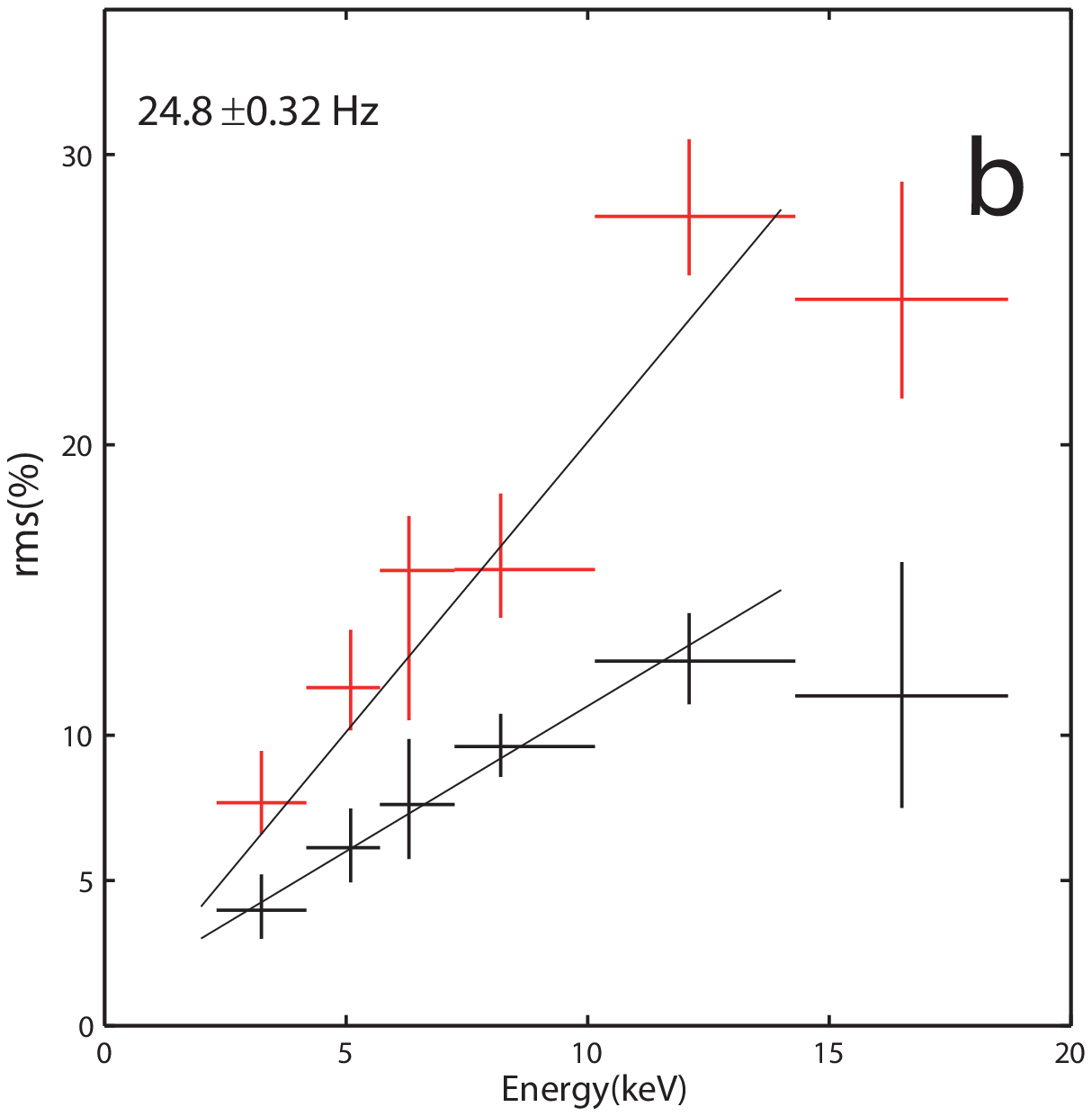}
\includegraphics[scale=.35]{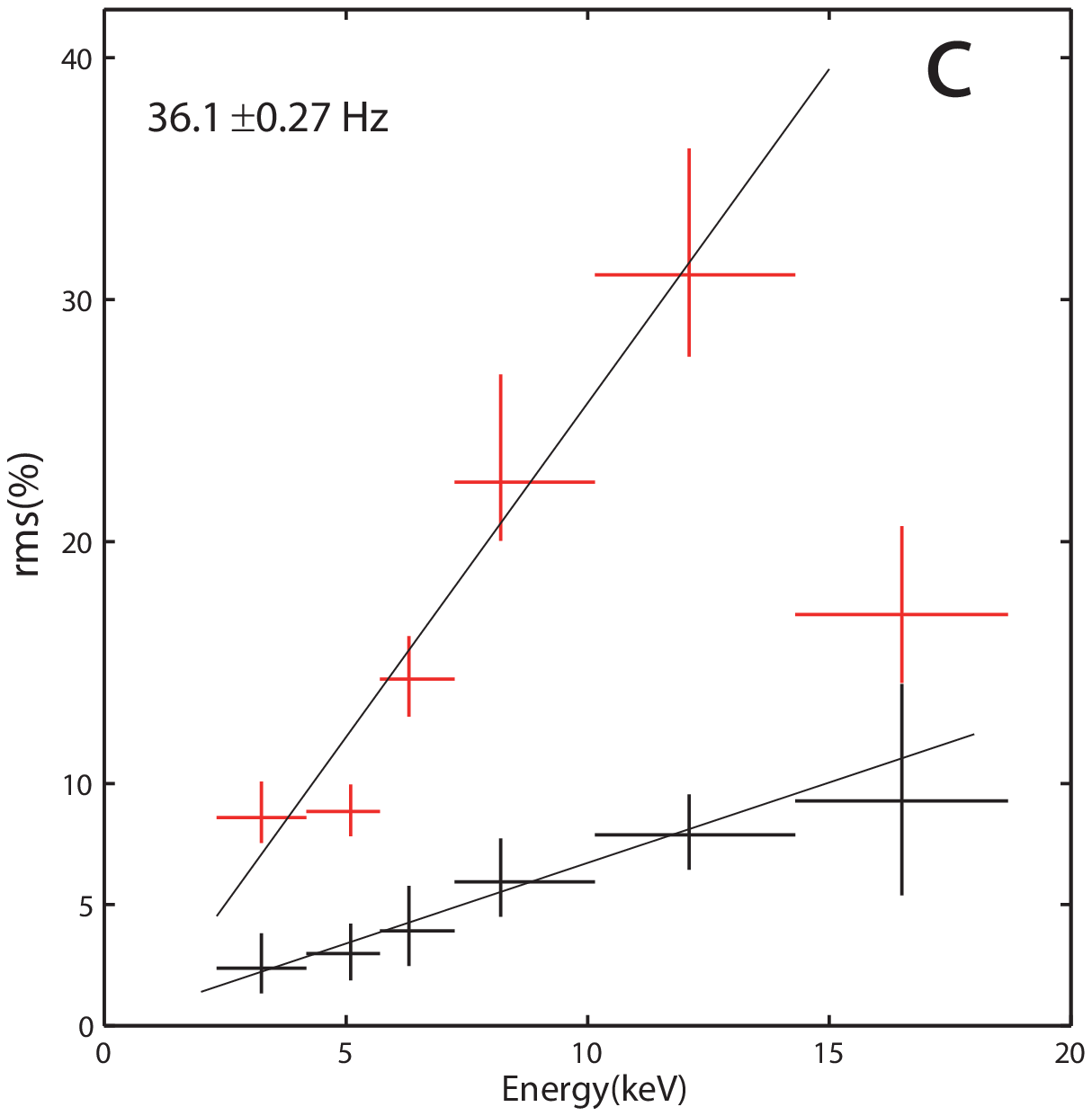}\\
\includegraphics[scale=.35]{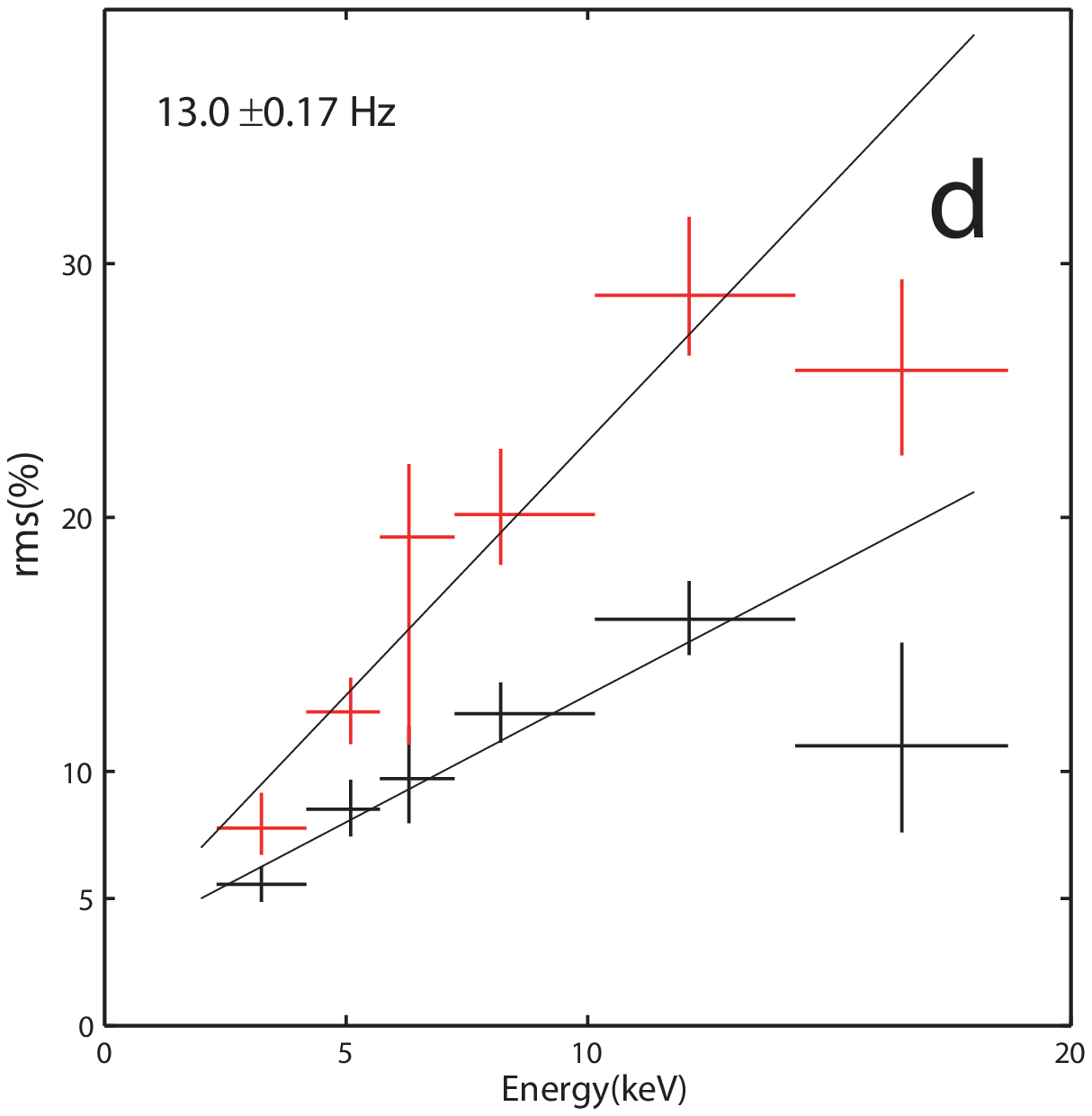}
\includegraphics[scale=.35]{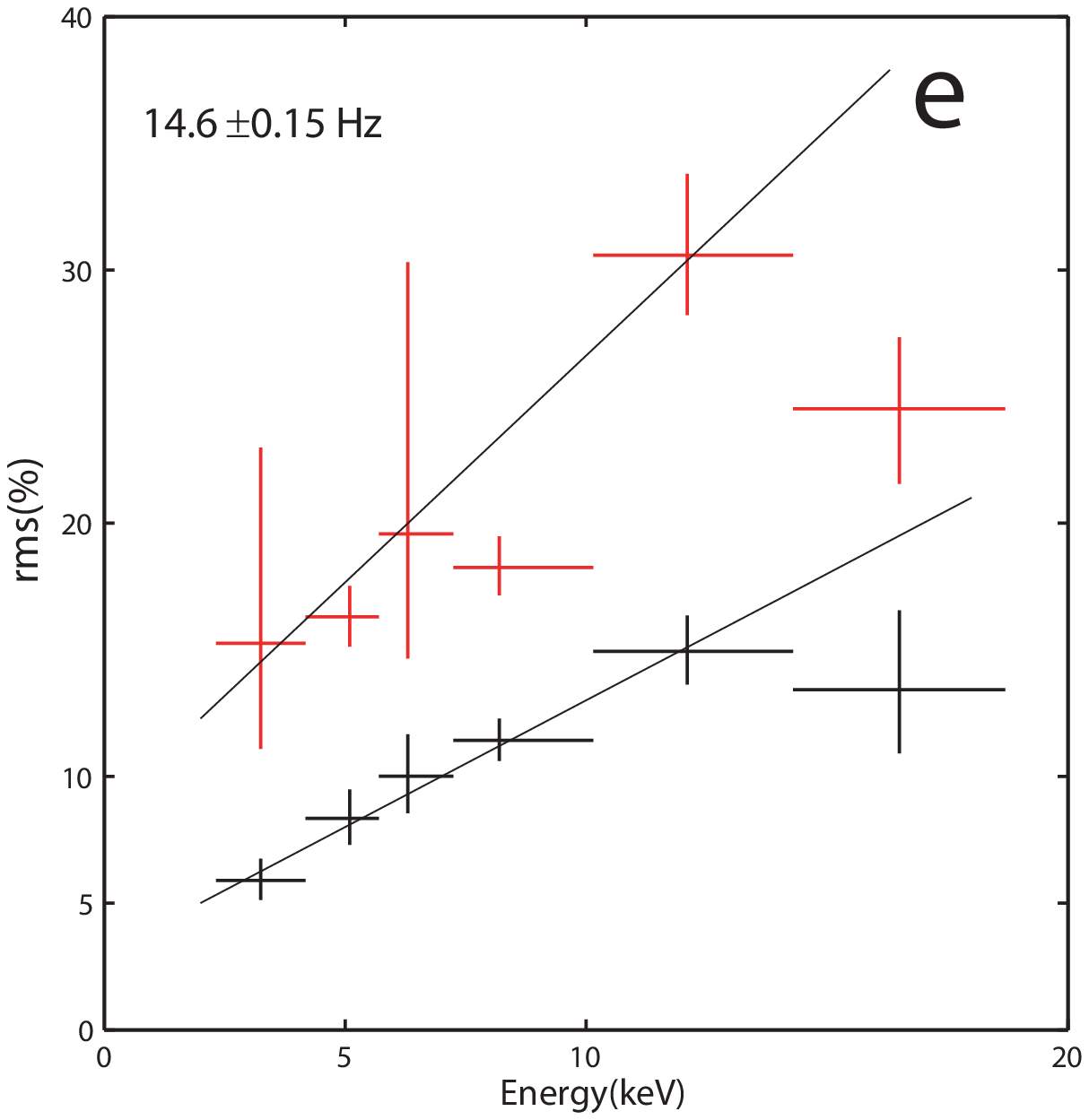}
\includegraphics[scale=.35]{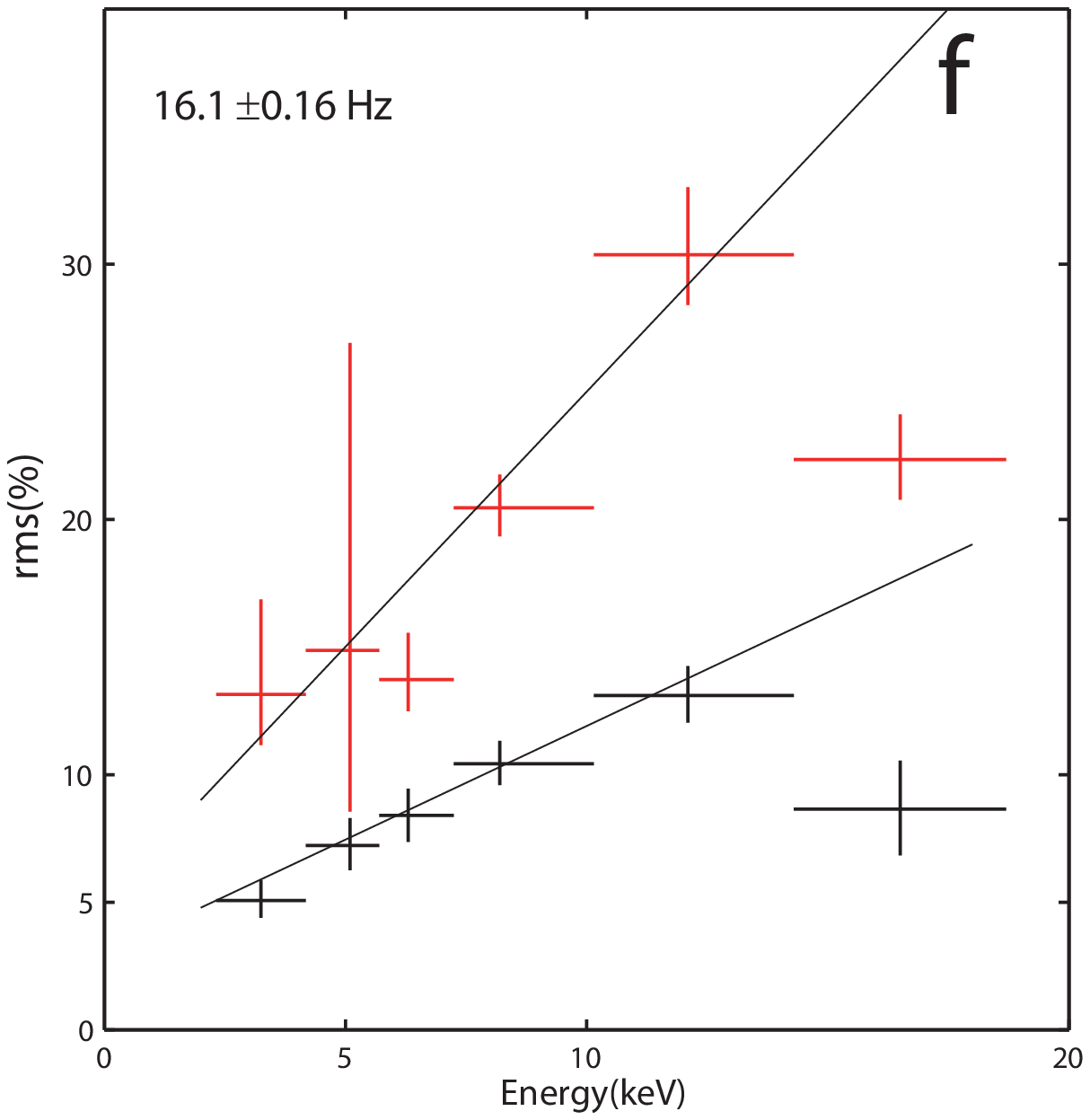}\\
\includegraphics[scale=.35]{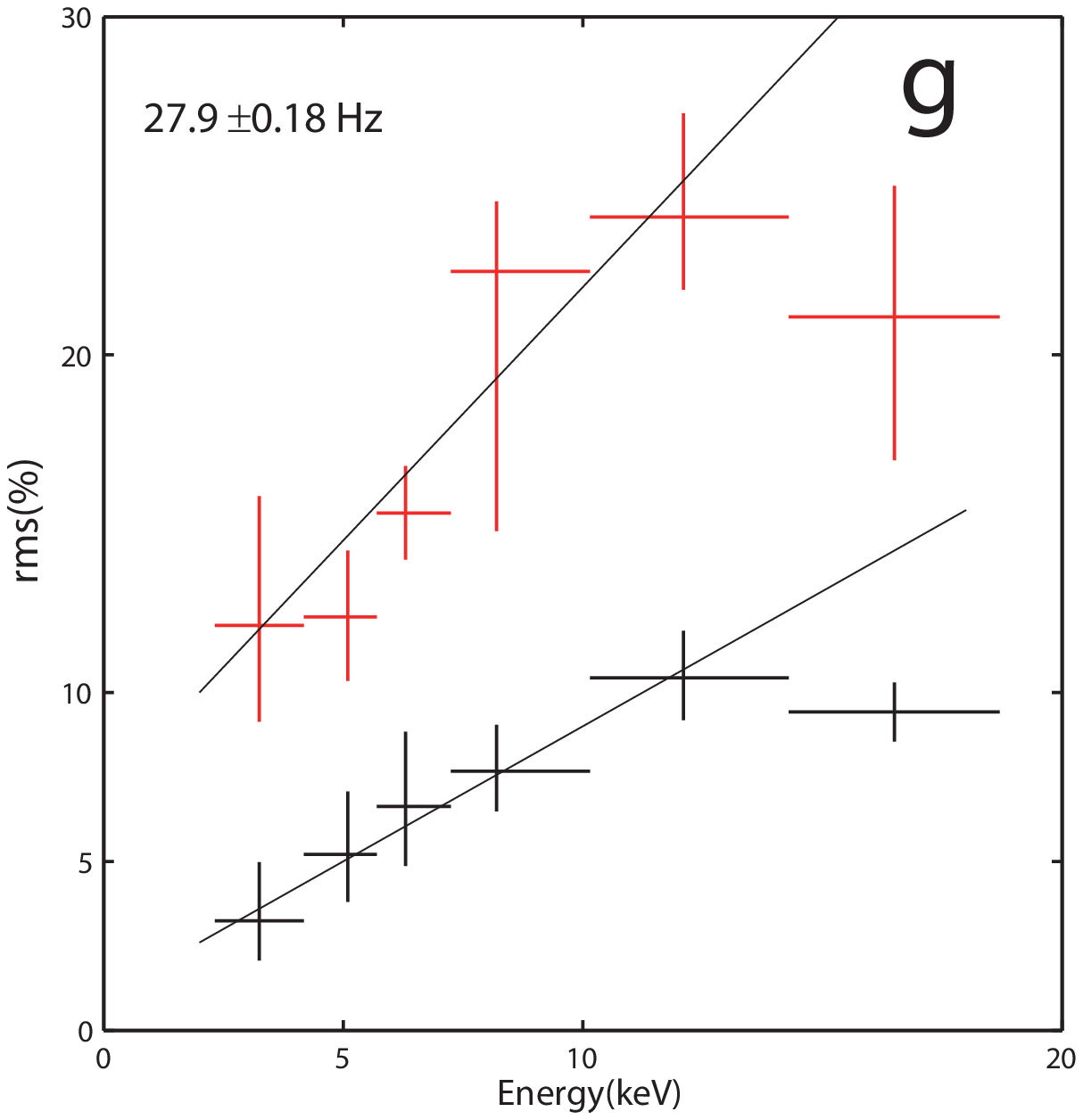}
\caption{This figure shows the rms-energy relation for seven selected boxes. (a)-(e) are extracted from the Cyg-like stage while (g) is extracted from the Sco-like stage of XTE J1701-462. The horizontal branch oscillation frequencies are marked in the upper left part of each sub-figure. Red crosses and black crosses represent $rms_{\rm{break}}$ and $rms_{\rm{HBO}}$, respectively.}
\label{fig5}
\end{figure}

\clearpage
\begin{figure}
\epsscale{1.1}
\plotone{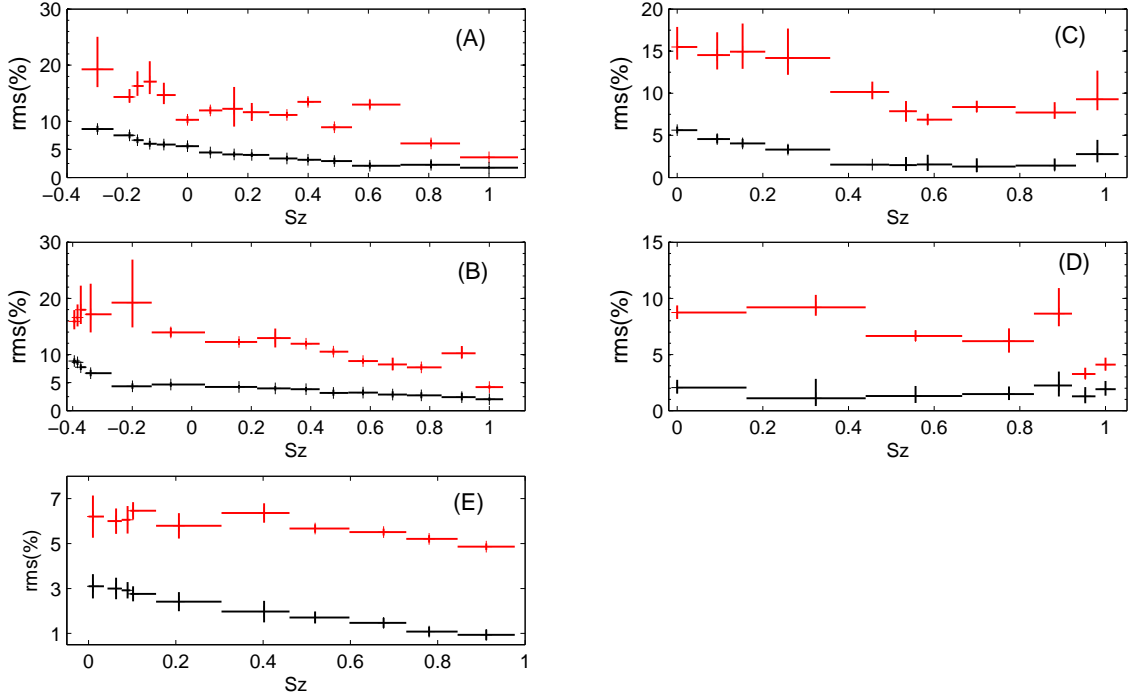}
\caption{ $rms-Sz$ relation of XTE J1701-462 and GX 17+2. (A)-(D) correspond, respectively, to intervals A-D, while (E) corresponds to interval E. Each subfigure shows the values of rms envolving from the (leftmost HB)upturn HB to HB/NB vertex. Red crosses and black crosses represent $rms_{\rm{break}}$ and $rms_{\rm{HBO}}$, respectively }
\label{fig6}
\end{figure}

\clearpage
\begin{figure}
\includegraphics[scale=.8]{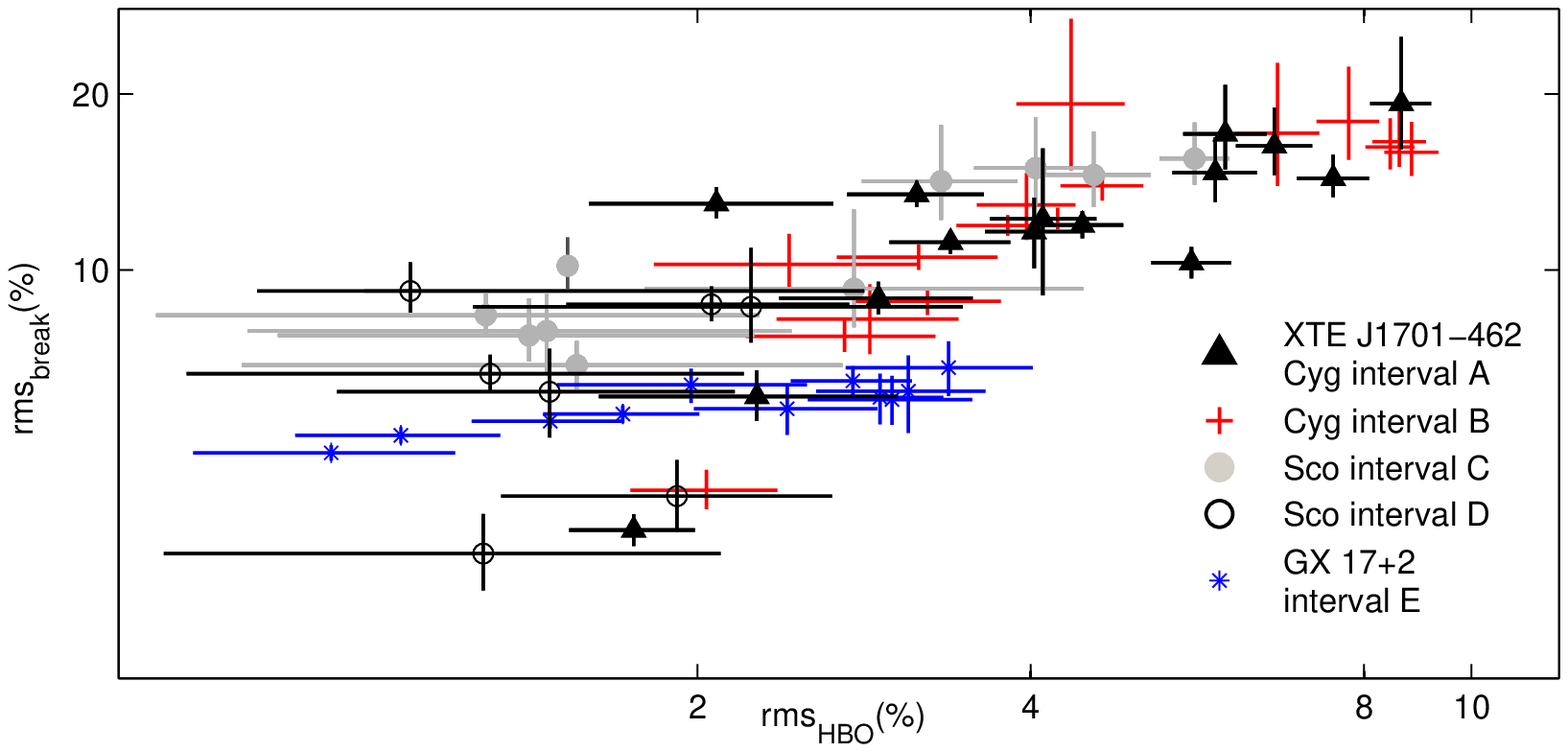} \\
\includegraphics[scale=.8]{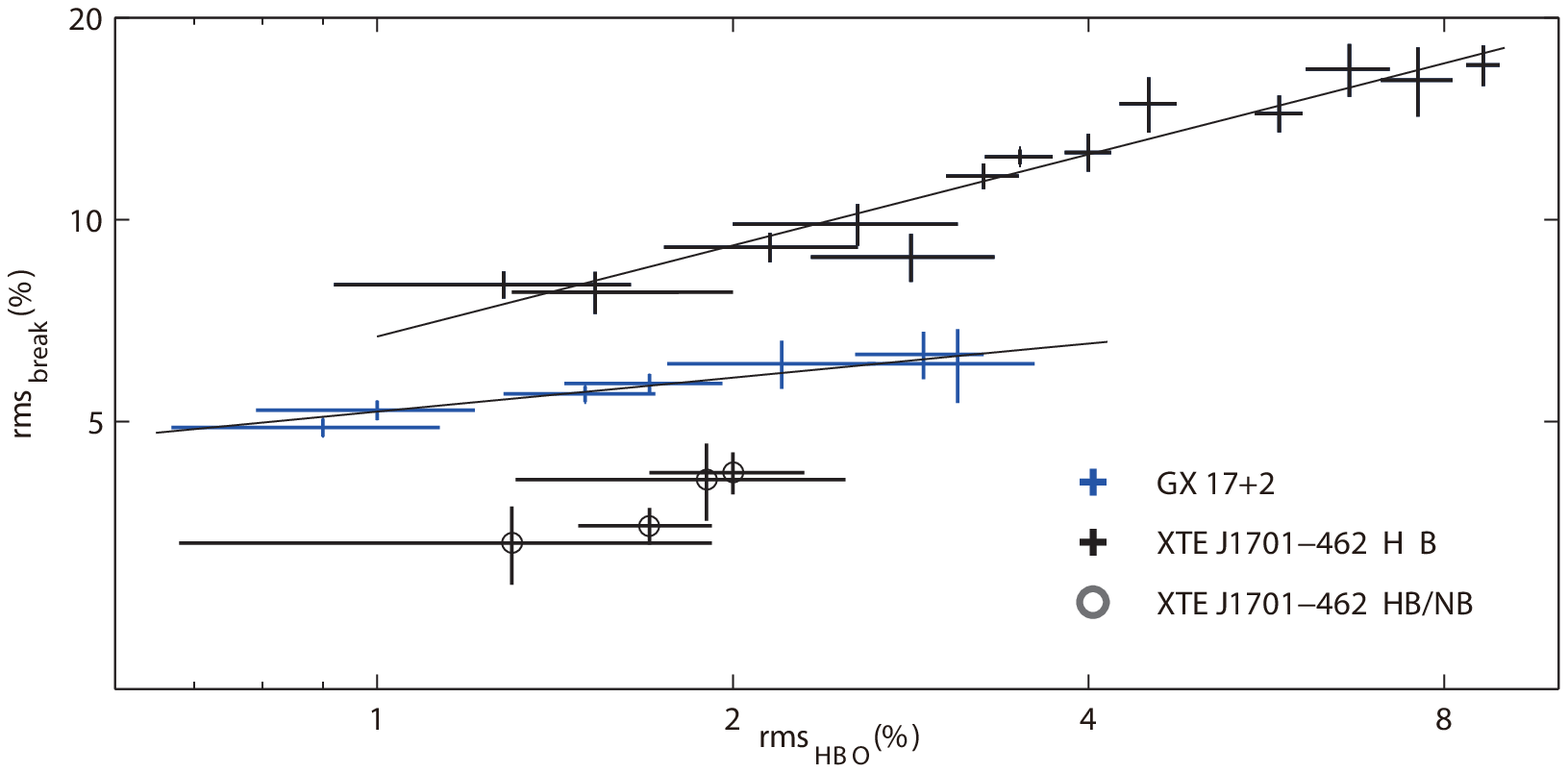}
\caption{ $rms_{\rm{break}}$-$rms_{\rm{HBO}}$ relation for XTE J1701-462 and GX 17+2. Figure 7b is a rebinned version of Figure 7a. The data in ellipse are extracted from HB/NB vertexes of XTE J1701-462.}
\label{fig7}
\end{figure}








\clearpage
\begin{deluxetable}{cccccc}
\tabletypesize{\scriptsize}
\tablecaption{Data selections for the extraction of hardness-intensity diagrams from XTE J1701-462 and GX 17+2. \label{tbl-1}}
\tablewidth{0pt}
\tablehead{
\colhead{Interval} & \colhead{Begin of date\ (DD/MM/YY)} & \colhead{Begin of Obs. } & \colhead{End of Date\ (DD/MM/YY)} & \colhead{End of Obs} & \colhead{Source type}}
\startdata
XTE J1701-462  \\ \hline
A& 21/01/06& 91106-01-04-00& 31/01/06& 91106-02-02-10& Cyg-like   \\
B& 02/02/06& 91106-02-02-14& 12/02/06& 91442-01-01-02& Cyg-like   \\
C& 17/02/06& 91442-01-07-02& 26/02/06& 91442-01-03-05& Sco-like   \\
D& 15/03/06& 92405-01-02-05& 23/03/06& 92405-01-03-06& Sco-like   \\ \hline
GX 17+2 \\ \hline
E& 03/10/99& 40018-01-02-000& 12/10/99& 40018-01-01-18& Sco-like  \\
\enddata
\end{deluxetable}

\clearpage

\begin{deluxetable}{cccccc}
\tabletypesize{\scriptsize}
\tablecaption{RXTE/PCA energy bands for XTE J1701-462 used in this work\label{tbl-2}}
\tablewidth{0pt}
\tablehead{
\colhead{PCA\ channel} & \colhead{Energy Band\ (keV)} & \colhead{Centroid Energy\ (keV)}}
\startdata
0-10& 2-4.5& 3.25   \\
11-13& 4.5-5.7& 5.1  \\
14-16& 5.7-6.9& 6.3  \\
17-22& 6.9-9.4& 8.2   \\
23-35& 9.4-14.8& 12.1  \\
36-149& 14.8-65& 16.5  \\
\enddata
\end{deluxetable}

\clearpage
\begin{deluxetable}{cccccc}
\tabletypesize{\scriptsize}
\tablecaption{Results from a multi-Lorentzian fitting for each box in Fig. 1. The parameters of both band-limited and HBO component are shown here. \label{tbl-3}}
\tablewidth{0pt}
\tablehead{
\colhead{box number} &\colhead{Sz} & \colhead{\ $\nu_{\rm{break}}$\ (Hz)} & \colhead{ $rms_{\rm{break}}$\ (\%)} & \colhead{$\nu_{\rm{HBO}}$\ (Hz)} & \colhead{$rms_{\rm{HBO}}$\ (\%)}}
\startdata
XTE J1701-462   Cyg-like interval A\\ \hline
1& -0.30& 2.8$\pm$0.1&  19.3$\pm$3.2&  12.1$\pm$0.1& 8.6$\pm$0.5 \\
4& -0.19& 3.5$\pm$0.1&  14.3$\pm$1.0&  22.9$\pm$0.4& 7.5$\pm$0.6 \\
5& -0.17& 4.0$\pm$0.0&  16.3$\pm$1.8&  24.7$\pm$0.3& 6.6$\pm$0.5 \\
6& -0.13& 4.3$\pm$0.0&  17.1$\pm$2.2&  26.8$\pm$0.3& 6.0$\pm$0.5 \\
7& -0.08& 4.6$\pm$0.2&  14.7$\pm$1.6&  28.4$\pm$0.3& 6.0$\pm$0.5 \\
8& 0& 6.3$\pm$0.5&  10.3$\pm$0.6&  31.7$\pm$0.5& 5.6$\pm$0.5 \\
9& 0.07& 6.6$\pm$0.4&  11.9$\pm$0.6&  35.3$\pm$0.2& 4.4$\pm$0.4 \\
10& 0.15& 6.9$\pm$1.1&  12.2$\pm$3.2&  37.0$\pm$0.3& 4.1$\pm$0.4 \\
11& 0.21& 6.9$\pm$0.6&  11.6$\pm$1.6&  39.3$\pm$0.4& 4.0$\pm$0.4 \\
12& 0.33& 7.7$\pm$0.7&  11.1$\pm$0.5&  41.0$\pm$0.5& 3.4$\pm$0.4 \\
13& 0.40& 7.3$\pm$0.3&  13.5$\pm$0.7&  41.9$\pm$0.6& 3.2$\pm$0.4 \\
14& 0.49& 9.9$\pm$0.9&  8.9$\pm$0.6&  43.2$\pm$1.0& 2.9$\pm$0.5 \\
15& 0.60& 9.2$\pm$0.2&  13.0$\pm$0.7&  46.8$\pm$1.2& 2.1$\pm$0.5 \\
16& 0.81& 12.1$\pm$1.6&  6.1$\pm$0.6&  48.7$\pm$1.2& 2.2$\pm$0.6 \\
17& 1& 8.6$\pm$0.6&  3.5$\pm$0.2& 51.6$\pm$0.7& 1.7$\pm$0.2 \\ \hline
XTE J1701-462   Cyg-like interval B\\ \hline
1& -0.40&  	2.2$\pm$0.0& 15.9$\pm$1.4& 13.0$\pm$0.2& 8.8$\pm$0.5 \\
2& -0.38&	2.5$\pm$0.0& 16.6$\pm$1.6&	14.3$\pm$0.2& 8.6$\pm$0.5 \\
3& -0.38&	2.5$\pm$0.0& 16.2$\pm$1.4&	14.9$\pm$0.2& 8.5$\pm$0.4 \\
4& -0.37&	2.9$\pm$0.1& 18.0$\pm$2.5&    15.4$\pm$0.1& 7.7$\pm$0.5 \\
5& -0.34&	3.3$\pm$0.2& 17.2$\pm$3.2&	16.7$\pm$0.2& 6.7$\pm$0.6 \\
6& -0.20&	4.5$\pm$0.1& 19.2$\pm$4.4&	33.0$\pm$0.3& 4.4$\pm$0.5 \\
7& -0.07&	5.6$\pm$0.3& 13.9$\pm$0.8&	34.2$\pm$0.2& 4.6$\pm$0.4 \\
8& 0.16& 	6.6$\pm$0.4& 12.2$\pm$0.5&	37.1$\pm$0.3& 4.2$\pm$0.3 \\
9& 0.28& 	6.8$\pm$0.5& 12.9$\pm$1.7&	39.1$\pm$0.5& 3.9$\pm$0.4 \\
10& 0.38& 	7.7$\pm$0.4& 11.9$\pm$0.5&	41.2$\pm$0.6& 3.8$\pm$0.4 \\
11 &0.48& 	8.7$\pm$	0.6& 10.5$\pm$0.5&	42.1$\pm$1.2& 3.2$\pm$0.5 \\
12& 0.58& 	10.0$\pm$0.8& 8.8$\pm$0.5&	43.1$\pm$1.3& 3.2$\pm$0.5 \\
13& 0.68& 	11.1$\pm$1.4& 8.2$\pm$1.1& 44.4$\pm$1.2& 2.8$\pm$0.5 \\
14& 0.77&   11.4$\pm$1.0& 7.7$\pm$0.5&	45.2$\pm$1.1& 2.7$\pm$0.5 \\
16& 0.91& 	11.5$\pm$0.5& 10.2$\pm$0.9& 48.2$\pm$2.1& 2.4$\pm$0.6 \\
17& 1&    10.2$\pm$1.1& 4.2$\pm$0.3&	51.3$\pm$0.9& 2.0$\pm$0.3 \\ \hline
XTE J1701-462   Sco-like interval C \\ \hline
1& 0&	6.0$\pm$	0.1& 15.5$\pm$1.5& 27.9$\pm$0.2& 5.6$\pm$0.4 \\
2& 0.09&  6.2$\pm$0.04& 14.5$\pm$1.7& 32.3$\pm$0.3& 4.6$\pm$0.5 \\
3& 0.15&  6.4$\pm$0.1& 14.9$\pm$2.1& 34.2$\pm$0.3& 4.0$\pm$0.5 \\
4& 0.26&	  6.6$\pm$0.1& 14.2$\pm$2.0& 35.9$\pm$0.4& 3.3$\pm$0.5 \\
5& 0.46&	  8.8$\pm$0.6& 10.2$\pm$0.9& 43.6$\pm$0.8& 1.5$\pm$2.5 \\
6& 0.53&  8.9$\pm$1.3& 7.9$\pm$1.3& 45.0$\pm$1.1& 1.5$\pm$0.7 \\
7& 0.59&  9.6$\pm$1.3& 6.9$\pm$0.6& 46.5$\pm$1.6& 1.6$\pm$0.8 \\
8& 0.70&	 9.8$\pm$0.6& 8.4$\pm$0.6& 48.9$\pm$1.3& 1.3$\pm$0.6 \\
9& 0.88&	 9.7$\pm$0.5& 7.7$\pm$0.8& 53.3$\pm$1.6	& 1.4$\pm$0.6 \\
10& 0.98&	11.1$\pm$0.5& 9.3$\pm$1.3& 	56.8$\pm$2.5& 2.8$\pm$1.0 \\ \hline
XTE J1701-462   Sco-like interval D\\ \hline
1& 0&	8.7$\pm$0.7& 8.7$\pm$0.6& 43.2$\pm$1.0& 2.1$\pm$0.5 \\
2& 0.32& 10.2$\pm$0.4&9.2$\pm$0.8& 47.1$\pm$1.5& 1.1$\pm$0.7 \\
3& 0.56& 9.6$\pm$1.1& 6.7$\pm$0.5& 46.2$\pm$1.2& 1.3$\pm$0.6 \\
4& 0.78& 10.1$\pm$1.7& 6.2$\pm$1.0& 52.9$\pm$1.4& 1.5$\pm$0.5 \\
5& 0.89& 8.7$\pm$0.3&	8.6$\pm$1.1& 56.6$\pm$3.4& 2.2$\pm$1.0 \\
6& 0.95& 8.7$\pm$1.5&	3.3$\pm$0.4& 58.2$\pm$1.8&	1.3$\pm$0.6 \\
7& 1& 9.7$\pm$1.7& 4.1$\pm$0.5& 60.7$\pm$1.6&	1.9$\pm$0.6 \\ \hline
GX 17+2  interval E\\ \hline
1& 0.01& 2.4$\pm$0.5& 6.2$\pm$1.0& 23.3$\pm$0.7& 3.1$\pm$0.5 \\
2& 0.06& 2.7$\pm$0.3& 6.0$\pm$0.6& 26.3$\pm$0.5& 3.0$\pm$0.5 \\
3& 0.09& 2.8$\pm$0.4& 6.1$\pm$0.6& 27.4$\pm$0.4& 2.9$\pm$0.4 \\
4& 0.10& 3.2$\pm$0.3& 6.5$\pm$0.4& 28.5$\pm$0.3& 2.8$\pm$0.3 \\
5& 0.21& 3.0$\pm$0.4& 5.8$\pm$0.6& 29.4$\pm$0.4& 2.4$\pm$0.4 \\
6& 0.40& 4.1$\pm$0.5& 6.4$\pm$0.4& 35.3$\pm$1.1& 2.0$\pm$0.5 \\
7& 0.52& 4.2$\pm$0.2& 5.7$\pm$0.2& 37.8$\pm$0.5& 1.7$\pm$0.3 \\
8& 0.68& 4.5$\pm$0.2& 5.5$\pm$0.2& 39.3$\pm$0.6& 1.5$\pm$0.2 \\
9& 0.78& 5.0$\pm$0.3& 5.2$\pm$0.2& 42.0$\pm$0.4& 1.0$\pm$0.2 \\
10&0.91& 5.4$\pm$0.3& 4.9$\pm$0.2& 44.2$\pm$0.6& 0.9$\pm$0.2 \\
\enddata
\end{deluxetable}

\clearpage
\begin{deluxetable}{ccccc}
\tabletypesize{\scriptsize}
\tablecaption{Data we used to study the PBK correlation in XTE J1701-462 and GX 17+2.\label{tbl-4}}
\tablewidth{0pt}
\tablehead{
\colhead{Observation(Sz)\tablenotemark{a}} & \colhead{$\nu_{\rm{HBO}}$\ (Hz)} & \colhead{$\nu_{\rm{low}}$\ (Hz)}}
\startdata
XTE J1701-462 \\ \hline
91442-01-07-09   &55.3$\pm$2.5  &642.7$\pm$3.1   \\
92405-01-01-02   &38.1$\pm$0.4  &615.1$\pm$3.8   \\
92405-01-01-04   &36.0$\pm$0.4  &502.4$\pm$23.1  \\
92405-01-02-03   &50.3$\pm$1.3  &623.9$\pm$8.3   \\
92405-01-02-05   &45.5$\pm$0.8  &595.9$\pm$10.1  \\
92405-01-03-05   &41.0$\pm$3.3  &612.5$\pm$7.7  \\
92405-01-40-04   &26.1$\pm$0.2  &650.5$\pm$7.1  \\ \hline
GX 17+2\\ \hline
0.80-0.88  &42.3$\pm$0.5  &518.2$\pm$6  \\
0.88-0.96  &45.1$\pm$0.7  &552.6$\pm$20.3 \\
1.00-1.04  &52.3$\pm$0.9  &619.5$\pm$22.9 \\
1.04-1.12  &53.3$\pm$1.7  &653.6$\pm$12.2 \\
1.12-1.20  &57.2$\pm$0.9  &690.3$\pm$16.1 \\
\enddata
\tablenotetext{a}{Notice: The data we used in this table are all extracted from \citet{san10} and \citet{lin12}. ObsID and Sz were used separately in their work.}
\end{deluxetable}


\clearpage
\begin{table}[h] \scriptsize
\caption{The parameters used to fit the rms-energy correlation of two componnets in Figure 5. The parameters with $1\sigma$ errors are given below, with R-squared stands for the coefficient of determination.\label{tbl-5}}
\begin{tabular}{*{7}{c}}
\hline
\multirow{2}*{figure number} &\multicolumn{3}{c}{break} &\multicolumn{3}{c}{HBO} \\
\cline{2-7}
 &a1 &b1 & R1-squared &a2 &b2 & R2-squared \\
\hline
\cline{1-7}
a &0.02$\pm$0.003 &0.04$\pm$0.02 &0.97  &0.01$\pm$0.001 &0.03$\pm$0.001 &0.99 \\
b &0.02$\pm$0.003 &0.001$\pm$0.03 &0.95 &0.01$\pm$0.001 &0.01$\pm$0.006 &0.99 \\
c &0.03$\pm$0.004 &-0.01$\pm$0.03 &0.97 &0.01$\pm$0.001 &0.00$\pm$0.006 &0.97 \\
d &0.02$\pm$0.004 &0.03$\pm$0.03  &0.94 &0.01$\pm$0.001 &0.03$\pm$0.001 &0.99 \\
e &0.02$\pm$0.003 &0.09$\pm$0.02 &0.93 &0.01$\pm$0.001 &0.03$\pm$0.007 &0.98 \\
f &0.02$\pm$0.004 &0.05$\pm$0.02 &0.93 &0.01$\pm$0.001 &0.03$\pm$0.006 &0.99 \\
g &0.015$\pm$0.004 &0.07$\pm$0.03 &0.90 &0.01$\pm$0.001 &0.01$\pm$0.006 &0.98 \\
\hline
\end{tabular}
\end{table}

\end{document}